\documentclass[12pt]{article}
\usepackage{amsmath}
 \usepackage{bbm}
\usepackage{amsfonts}
\usepackage{cite}
\newsavebox{\myhbar}
\savebox{\myhbar}{$\hbar$}
\usepackage{graphicx}
\usepackage{enumerate}
\usepackage{mdframed}
\usepackage{url}
\usepackage{wasysym}
\usepackage[hmargin=2.4cm,vmargin=2.8cm]{geometry}
\usepackage{undertilde}
\usepackage{slashed}
\numberwithin{equation}{section}
\renewcommand*{\hbar}{\mathalpha{\usebox{\myhbar}}}
\begin{document}
\title{Decays of unstable quantum systems}
\author{Charis Anastopoulos\footnote{anastop@physics.upatras.gr}\\ {\small Department of Physics, University of Patras, 26500 Greece} }
\maketitle

\begin{abstract}
This paper is a  pedagogical yet  critical  introduction to the quantum description of  unstable systems,  mostly at the level of  a graduate quantum mechanics course. Quantum decays appear in many different fields of physics, and their description beyond the exponential approximation is the source of technical and conceptual challenges.
In this article, we present both general methods that can be adapted to a large class of problems, and specific elementary models to describe phenomena
 like   photo-emission, beta emission and tunneling-induced decays. We pay particular attention to the emergence of exponential decay;  we analyze   the approximations that justify it,  and we present criteria for its breakdown. We also present  a detailed model for non-exponential decays due to resonance, and an elementary model describing  decays in terms of particle-detection probabilities.  We argue that the traditional methods for treating decays face significant problems outside the regime of exponential decay, and that the exploration of  novel regimes of current interest requires new tools.

\end{abstract}

\section{Introduction}

\subsection{Key notions}

A decay of a particle $A$ is a process of the type
 $A \rightarrow B_1 + B_2 + \ldots B_n$, where  the particles  $B_i$ are  the decay products. Decays are ubiquitous in physics. Examples include the emission of photons from excited atoms or nuclei, alpha and beta emission, decays of composite subatomic particles (for example, neutrons or  pions) and decays of elementary particles (for example, a muon decaying to one electron and two neutrinos).

Most decays follow an exponential law.  The probability that a decay takes place within the time-interval
  $[t, t+ \delta t]$, for $t > 0 $,  equals  $p(t) \delta t$, where the probability density $p(t)$ is given by
\begin{eqnarray}
p(t) = \Gamma e^{- \Gamma t}. \label{probdec}
\end{eqnarray}
The {\em decay constant} $\Gamma$ is positive and has dimensions of inverse time.

We note that for any two instants of time $t_1$ and $t_2$,
\begin{eqnarray}
p(t_2) = p(t_1) e^{-\Gamma(t_2-t_1)},
\end{eqnarray}
Hence, the exponential law remains invariant under a shift of the     initial moment of time $t = 0$. The decay of an ensemble of $A$ particles after a moment of time $t$ carries no memory of any properties prion to $t$. In classical  probability theory, exponential decays correspond to {\em Markovian processes}.

Experiments typically involve a large number of decaying particles with identical preparation and the detection of  some of  the decay products. Recording the number of detection events within given time intervals $[t, t+\delta t]$, we can reconstruct the probability density $p(t)$ associated to the decay. Hence,  $p(t)$ is a {\em directly observable quantity}.

The focus of this paper is the derivation of the probability density $p(t)$ from the rules of quantum mechanics. We present different approaches to the problem, we apply them to specific physical problems and we analyze their underlying assumptions and their limitations.

\subsection{The quantum description of decays}

The aim of the quantum mechanical description of a decay process is to construct the probability density $p(t)$ from first principles. This probability density is different from the ones usually considered in quantum theory, because  the random variable $t$ is   temporal.  We cannot use Born's rule, because there is no self-adjoint operator for time in quantum theory \cite{Pauli}. Indeed, the construction of quantum probabilities in which time is a random variable is an old problem---for reviews, see,  Ref.   \cite{ToAbooks}. There are several different approaches that lead to different results, even for elementary problems, for example, constructing probabilities for the time of arrival \cite{ML} or   specifying the time it takes a particle to tunnel through a potential barrier \cite{tunt}.

 \medskip
 {\em The persistence amplitude method.}
Most studies of decays avoid  a direct construction of $p(t)$.  Instead they focus on a slightly different issue,  namely, on finding the probability that the quantum system persists in its initial  configuration. In quantum theory, the notion of a configuration refers to the set of all quantum states compatible with a specific property. Mathematically, it corresponds  to a subspace of the system's Hilbert space, and it is represented by the associated projection operator. For example, if the defining property of the initial configuration is that a particle is confined by a potential well in a spatial region $U$, the relevant subspace corresponds to the projection operator $\hat{P}_U = \int_U dx |x\rangle \langle x|$ that describes the property $"x \in U"$ for the particle position $x$.

 In some unstable systems, the subspace of the initial configuration is one-dimensional, and it coincides with the quantum state  $|\psi\rangle$ at which the system has been initially prepared. For such systems,
it is convenient to employ the
   {\em persistence amplitude} (or {\em survival amplitude})

\begin{eqnarray}
{\cal A}_{\psi}(t) = \langle \psi|e^{-i\hat{H}t}|\psi\rangle. \label{persiss}
\end{eqnarray}
where $\hat{H}$ is the Hamiltonian of the system---for the mathematical properties of the persistence amplitude, see Refs. \cite{GhiFo, Peres80}.

The modulus square of
  ${\cal A}_{\psi}(t)$ is the {\em persistence probability} (or {\em survival probability}), i.e., the probability  a system prepared at the state $|\psi\rangle$ at $t = 0$ will still  be found at $|\psi\rangle$ by a measurement at a later time $t$. The persistence probability is identical with the {\em fidelity} between the initial state and its time-evolution.

   It is then suggested that the decay probability density be defined as
\begin{eqnarray}
p(t) = - \frac{d}{dt} |{\cal A}_{\psi}(t)|^2, \label{decprob2}
\end{eqnarray}
i.e., the decay is assumed to happen when the system leaves  $|\psi\rangle$.

Eq. (\ref{decprob2}) is a reasonable candidate probability density provided that (i) all states  normal to $|\psi\rangle$ that can be reached via Hamiltonian evolution correspond to decay products, and (ii) the probability of the reverse  process to the decay is negligible.

If condition (i) is not satisfied, then the propositions A = "the system left $|\psi\rangle$" and B = "the decay happened" are not identical: B implies A, but A does not imply B. Hence,  Eq. (\ref{decprob2}) cannot be identified with the decay probability. For example,  if
 $|\psi\rangle$ evolves to a different state $|\phi\rangle$ that describes the initial particle $A$,  a part of the persistence amplitude will describe  Rabi-type oscillations between $|\psi\rangle $ and $|\phi \rangle$. The persistence probability may not be a decreasing function of $t$, and, hence, $p(t)$ may take negative values. It will not be a genuine probability density: Eq. (\ref{decprob2}) will not predict the number of particles detected at each  moment of time.

Condition (ii) is satisfied if there are many more states available to the decay products than to the initial particle. Furthermore, the configuration of the experiment must be such as to allow the decay products to `explore' their available states. This means that the decay products  leave the locus of their production and they are measured far away. Consider for example an excited atom contained in a cavity. For some cavity geometries,  the emitted photon does not exit the cavity immediately, and it may be reabsorbed by the atom at a later time. The persistence amplitude of the excited atomic state may has an oscillating component. In this case, the candidate probability density (\ref{decprob2}) does not correlate with  photodetection records.

For systems that satisfy conditions (i) and (ii), Eq. (\ref{decprob2}) provides a  reasonable  construction of the decay probability density. It works best in model systems that incorporate the conditions (i) and (ii)  above into their definition. On such example is   the Lee model \cite{Lee} that is presented in   Sec. 3. However, even in such models the candidate probability density (\ref{persiss}) may become negative outside the exponential decay regime---see, Sec. 3.1.3.

We note that the interpretation of the persistence probability as the fidelity of the initial state makes it a useful tool also for the study of a broader class of physical phenomena than decays \cite{BPSZ}. Hence, the quantity (\ref{decprob2}) may be of physical interest, even if it takes negative values and cannot be identified with the probability density for decay.
 \medskip

 {\em Probability currents.}
 An alternative elementary description of decays is available, whenever we can  associate a probability-current operator $\hat{\pmb J}({\pmb x}, t)$ to one of the decay products. For example, for non-relativistic particles of mass $m$ satisfying Schr\"odinger's equation with Hamiltonian $\hat{H}$, the current operator is
\begin{eqnarray}
 \hat{\pmb J}({\pmb x}, t) = \frac{1}{2m}e^{i\hat{H} t}\left[ \hat{\pmb p} \delta^3(\hat{\pmb x} - {\pmb x}) + \delta^3(\hat{ \pmb x} - {\pmb x}) \hat{\pmb p} \right] e^{-i\hat{H}t},
\end{eqnarray}
where $\hat{\pmb x}$ and $\hat{\pmb p}$ are the standard position and momentum operators, respectively.
The expectation value of  $\hat{\pmb J}({\pmb x}, t)$ on a   state $|\psi\rangle $ reproduces the standard expression for the probability current $\frac{1}{m}[\mbox{Im} \psi^*({\pmb x}, t) {\pmb \nabla} \psi({\pmb x}, t)]$  associated to the solution $\psi({\pmb x}, t) := \langle {\pmb x}|e^{-i \hat{H}t}|\psi\rangle$ of Schr\"odinger's equation.

Given a current operator  $\hat{\pmb J}({\pmb x}, t)$ for one of the decay products, we can evaluate the flux  $\Phi_C(t) = \int_C d^2{\pmb \sigma} \cdot \langle \psi| \hat{\pmb J}({\pmb x}, t)|\psi\rangle$ through any surface $C$. In many set-ups, the probability flux  coincides with the particle flux through this surface, which is  a directly measurable quantity. Then, the flux $\Phi_S(t)$ over a two-sphere surrounding the unstable system is expected to be proportional to the decay probability density $p(t)$.

 The main limitation of this method is that it cannot be consistently applied to relativistic systems, because of difficulties in defining probability currents that are both  Lorentz covariant and causal \cite{RoHo}. For this reason, it has mainly been used only in non-relativistic settings, in particular, for decays that can be described in terms of tunneling \cite{LL, Perel}.

 Another problem of the probability current method is the possibility of back-flow: the current operator has  negative eigenvalues even for states with strictly positive momentum \cite{Brack}. Hence, in some cases the probability flux may turn out to be negative for some times $t$--especially when the current is evaluated near the locus of the decay \cite{Winter}. In such cases the flux is not a reliable measure of detected outgoing particles.

 \medskip

 {\em Detector models.} A more rigorous description of quantum decays requires the incorporation of  the measurement process  \cite{EkSi, GhiFo}. In fact, the idea that quantum decays cannot be explained solely in terms of unitary time evolution was  crucial to the early interpretational discussions of quantum theory \cite{Heismem}. We have to take into account   the irreversibility of the quantum measurements on the decay products  in order to avoid theoretical discrepancies. This is the case, for example, in systems characterized by competing decay channels. If one ignores the effects of the apparatus, the persistence amplitude  exhibits large oscillations, thereby contradicting the standard classical description of sequential decays \cite{FoGhi}.

In quantum measurement theory, the consistent treatment of the measuring apparatus allows of the description of quantum observables in terms of Positive Operator Valued Measures (POVM), i.e., a set of positive operators $\hat{\Pi}(a)$, where $a$ are the values of the physical magnitude recorded by the apparatus. Then, given an initial state $\hat{\rho}$ of the system, the probability that the result $a$ is obtained is given by $Tr\left[\hat{\rho}_0 \hat{\Pi}(a)\right]$.

 In recent years, a new class of model for particle detection has been developed, allowing    for  the construction of quantum observables for the time $t$ of a detection event \cite{AnSav}---see, also \cite{An08} for an early application to the decay problem. In this paper, we present a study of decays, using one such
observable $\hat{\Pi}(t)$ for detection time  that can be obtained by  elementary arguments similar to the ones of standard photodetection theory \cite{Glauber}.

\subsection{This paper}
The aim of this paper is to provide the
 reader with tools for addressing a large class of decay problems. We present the most common approximation schemes, as well as criteria for checking when they fail. We prefer to work with models that admit simple solutions and controlled approximations, but we introduce more complex models for the explicit demonstration of important physical issues.

A key motivation for this paper is to provide explicit arguments that the traditional methods for describing decays (persistence amplitude and probability current) are inapplicable for many problems of current physical interest.  These methods work excellently for exponential decays, but they lead to negative values of the decay probability $p(t)$ outside the exponential regime. The problem is not a failure of some approximation,  the  very definition of such methods cannot guarantee positivity. We believe that a first-principles construction of decay probabilities is essential for describing decays in novel regimes, for example, in entangled systems \cite{AnHu2, CVS17, C18}, attosecond tunneling ionization \cite{atto} or decay oscillations in nuclear physics \cite{GSI}.

The paper is organized as follows.
  Sec. 2 presents the simplest method for the study of decays, namely, the evaluation of the persistence amplitude as a line integral on the complex energy space. This method is particularly suitable for perturbative decays, i.e., systems in which the decay originates from a small term in the Hamiltonian of the system. It also applies to relativistic systems described by quantum field theory.

  We show that   exponential decay is common in perturbative decays: it originates  from the separation of energy scales in the persistence amplitude. Nonetheless, it is neither exact nor generic. Exponential decay fails at very early and very late times, and also if the energy of the initial state is close to a resonance of the system.

 In  Sec. 3, we present the Lee model that describes the decaying particle as a  two-level system. This model can easily be adapted to problems in different branches of physics, including high energy physics, nuclear physics, atom optics, and condensed matter. Here, we present its applications to photo-emission and beta decay. We also employ the model in order to demonstrate the breakdown of the persistence amplitude method outside the exponential decay regime.

 Sec. 4 describes the decays of an atom in a cavity, providing an explicit example of decays that are non-exponential at all times because of resonance.

In Sec. 5, we study non-perturbative decays due to quantum tunneling, using the probability current method. We show that exponential decays originate from a decoherence condition, namely the lack of interferences between different attempts of the particle to tunnel through the barrier.

In Sec. 6, we revisit the Lee  model using probabilities constructed from a simple temporal observable, analogous to the one used in photodetection theory. We show that the detection probability reproduces the results of the persistence amplitude method, but can also be used in regimes where the latter fails.

The material in this article is mostly at the level of a graduate course in quantum mechanics. Familiarity with quantum many-particle systems is presupposed.
Secs. 2, 3 and  5 provide a minimal   introduction to quantum decays, explaining the emergence of the exponential decay law and its limitations,  developing methods that can be applied to different branches of physics, and presenting some important    physical examples.
The Appendix contains a sketch of  additional results that can be used as exercises.

\section{Perturbative evaluation of the persistence amplitude}

In this section, we study decays in which the persistence amplitude can be evaluated perturbatively. All physical properties of the decays are encoded into a function defined on the complex energy plane, the {\em self-energy function}. We identify the conditions under which the exponential law emerges, and we identify regimes for non-exponential decays.

\subsection{ Preliminaries}
By definition, the initial state
 $|\psi\rangle$ of an unstable system is not an eigenstate of the Hamiltonian $\hat{H}$. However, in many problems of physical interest, $|\psi\rangle$ is close to an eigenstate of $\hat{H}$, in the sense that  $\hat{H}$  is of the form
\begin{eqnarray}
\hat{H} = \hat{H}_0 + \hat{V}, \label{Hamper}
\end{eqnarray}
where $|\psi\rangle$ is an eigenstate of $\hat{H}_0$ and  $\hat{V}$ is a small perturbation. For example, $\hat{H}_0$ may be the Hamiltonian of an atom, and $\hat{V}$ the interaction Hamiltonian of the atom with the quantum electromagnetic field.
In this case, we evaluate the persistence amplitude   perturbatively.

Let us denote by $|b\rangle$ the eigenstates of $\hat{H}_0$ and by $E_b$ the corresponding eigenvalues. Then,
\begin{eqnarray}
\hat{H}_0 |b\rangle = E_b|b\rangle.
\end{eqnarray}
We will take the initial state $|\psi\rangle$ to be one of the eigenstates, say $|a\rangle$, and we will write  $V_{ab} := \langle a|\hat{V}|b\rangle$.

Without loss of generality, we     assume that  $\langle a|\hat{V}|a\rangle =0$. If $\langle a|\hat{V}|a\rangle \neq 0$,
  we can always  redefine
 \begin{eqnarray}
 \hat{H}_0' = \hat{H}_0 + \sum_a \langle a|\hat{V}|a\rangle |a\rangle \langle a| \hspace{1cm}  \hat{V}' = \hat{V} - \sum_a \langle a|\hat{V}|a\rangle |a\rangle \langle a|,
  \end{eqnarray}
so that  $|a\rangle$ is an eigenstate of $\hat{H}_0'$ and $\langle a|\hat{V}'|a\rangle = 0$.

\medskip

\noindent {\em The resolvent.} The perturbative calculation of the persistence amplitude is simplified by using the {\em resolvent} associated to the Hamiltonian operator:
  $(z - \hat{H})^{-1}$, for $z \in {\pmb C}$. We will often write the resolvent  as $\frac{1}{z - \hat{H}}$.

The resolvent is related to the evolution operator $e^{-i\hat{H}t}$ by

\begin{eqnarray}
e^{-i\hat{H}t} = \lim_{\epsilon \rightarrow 0} \frac{i}{2\pi  } \int_{-\infty }^{\infty } \frac{dE e^{-iEt}}{E+ i \epsilon - \hat{H}},  \label{propagator2}
\end{eqnarray}
where $\epsilon > 0$  and $t > 0$.

Eq. (\ref{propagator2}) follows from  the identity
\begin{eqnarray}
e^{- i \omega t} =  \lim_{\epsilon \rightarrow 0} \frac{i}{2\pi  } \int_{-\infty }^{\infty } \frac{dE e^{-iEt}}{E+ i \epsilon - \omega}. \label{idtycmplx}
\end{eqnarray}
 To prove Eq. (\ref{idtycmplx}) we evaluate the line integral

 \begin{eqnarray}
I(t) = \oint_{C_+} dz \frac{e^{-izt}}{z - \omega}, \label{cont}
\end{eqnarray}
along the negative-oriented contour $C_+$ of Fig. 1. At the lower half of the imaginary plane $Im z  =  - y$, for $y >0$. Therefore, the integrand along the semicircle is proportional to $e^{-yt}$ and vanishes as the radius of the semi-circle tends to infinity. Hence,
\begin{eqnarray}
I(t) = \lim_{\epsilon \rightarrow 0} \int_{-\infty }^{\infty } \frac{dE e^{-iEt}}{E+ i \epsilon - \omega}.
\end{eqnarray}
The contour $C_+$ includes the single pole of the integrand (\ref{cont}), at $z = \omega$. Using Cauchy's residue theorem, we arrive at Eq. (\ref{idtycmplx}).

By integrating along the contour $C_-$, we can similarly prove that
\begin{eqnarray}
\lim_{\epsilon \rightarrow 0}   \int_{-\infty }^{\infty } \frac{dE e^{-iEt}}{E -  i \epsilon - \hat{H}} = 0 ,  \label{propagator2b}
\end{eqnarray}
for $\epsilon > 0$  and $t > 0$. The   integral vanishes because the contour $C_-$ does not enclose any pole.

\begin{figure}[]
\includegraphics[height=6cm]{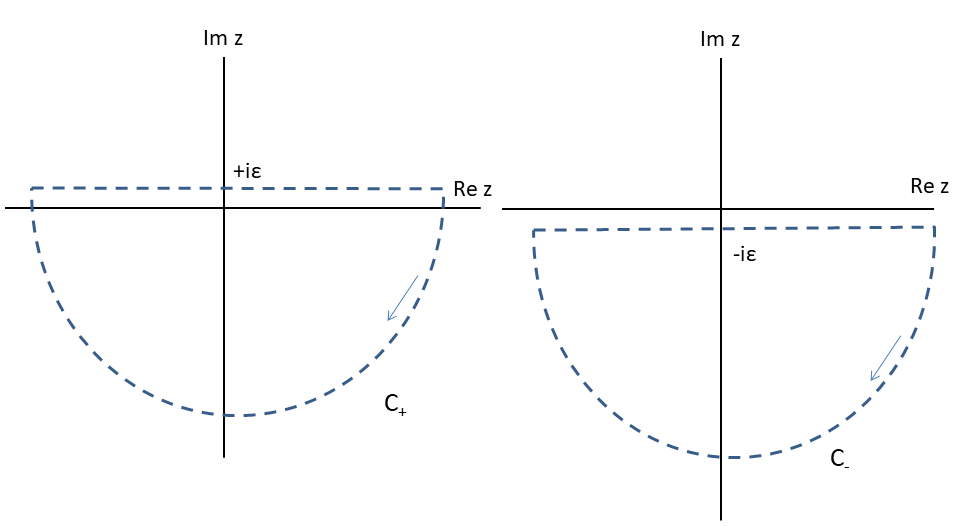} \caption{ \small Integration contours $C_{\pm}$ for calculating the integrals  (\ref{idtycmplx}) and (\ref{propagator2b}).  The integrals are evaluated at the limit where the radius $r$ of the semi-circle goes to infinity. Both curves are traversed clock-wise and therefore have negative orientation.}
\end{figure}

A crucial property of the resolvent is that it can be expanded in a perturbative series. For a Hamiltonian $\hat{H}$ of the form (\ref{Hamper}),
 \begin{eqnarray}
  (z - \hat{H})^{-1}
 = (z - \hat{H}_0 - \hat{V})^{-1} = [(z-\hat{H}_0) (1 - (z-\hat{H}_0)^{-1}\hat{V})]^{-1}\nonumber \\
  = (1 - (z-\hat{H}_0)^{-1}\hat{V})^{-1}(z-\hat{H}_0)^{-1}. \nonumber
 \end{eqnarray}
 Using the geometric series formula $(1 - \hat{A})^{-1} = \sum_{n=0}^{\infty} \hat{A}^n$, we obtain
\begin{eqnarray}
\frac{1}{z - \hat{H}} = \frac{1}{z - \hat{H}_0} + \frac{1}{z - \hat{H}_0}  \hat{V} \frac{1}{z - \hat{H}_0}  + \frac{1}{z - \hat{H}_0}  \hat{V}\frac{1}{z - \hat{H}_0}  \hat{V} \frac{1}{z - \hat{H}_0}  + \ldots \label{seriesprop}
\end{eqnarray}

\subsection{The random phase approximation}
We construct the persistence amplitude  ${\cal A}_a(t)$ for an initial state $|a\rangle$ that is an eigenstate of $\hat{H}_0$. By  Eq. (\ref{propagator2}),
 \begin{eqnarray}
 {\cal A}_a(t) = \lim_{\epsilon \rightarrow 0} \frac{i}{2\pi } \int_{-\infty }^{\infty } dE e^{-iEt}  G_a(E+i\epsilon) , \label{ampll}
 \end{eqnarray}
 where
  \begin{eqnarray}
  G_a(z) =  \langle a|(z - \hat{H})^{-1}|a\rangle.
 \end{eqnarray}

 We evaluate $G_a(z)$ using the perturbative series (\ref{seriesprop}). We assume that $\hat{V}$ is of first order to  some  dimensionless parameter $\lambda << 1$.  The zeroth-order contribution to $G_a(z)$ is
  $ \frac{1}{z - E_a}$. The first-order contribution is
 $ \frac{1}{(z - E_a)^2} \langle a|\hat{V}|a\rangle$ = 0.

 The second-order term is  $ \frac{1}{(z - E_a)^2} \langle a|\hat{V} \frac{1}{z - \hat{H}_0} \hat{V}|a\rangle$. Writing $ \frac{1}{z - \hat{H}_0} =  \sum_b \frac{1}{z - E_b} |b\rangle \langle b|$, this term becomes $ \frac{1}{(z - E_a)^2} \Sigma_a(z) $, where
\begin{eqnarray}
\Sigma_a(z) = \sum_b \frac{|V_{ab}|^2}{z - E_b}, \label{sigmaa}
\end{eqnarray}
is the {\em self-energy} function of the state $|a\rangle$.
 The third-order term is
 \begin{eqnarray}
\sum_{bc} \frac{1}{(z-E_a)^2} \frac{ V_{ab} V_{bc} V_{ca}}{(z-E_b)(z-E_c)}, \nonumber
\end{eqnarray}
and the fourth-order term is
\begin{eqnarray}
\sum_{bcd} \frac{1}{(z-E_a)^2} \frac{ V_{ab} V_{bc} V_{cd}V_{da}}{(z-E_b)(z-E_c)(z-E_d)}. \nonumber
\end{eqnarray}
 The procedure can be continued ad infinitum.

We assume that the Hamiltonian $\hat{H}_0$ has continuous spectrum for $E > \mu$, for some parameter $\mu$. The continuous spectrum corresponds to the kinetic energy of the decay products. With this assumption, we invoke the   {\em Random Phase Approximation} (RPA), according to which

\begin{eqnarray}
\sum_c \frac{V_{ac}V_{cb}}{z-E_c} \simeq \delta_{ab} \Sigma_a(z). \label{rpa}
\end{eqnarray}
The reasoning for Eq. (\ref{rpa}) is the following. Suppose that the system is contained in a box of volume $V$ with periodic boundary conditions and that it contains a large number $N$ of degrees of freedom. The RPA is the assumption that the phases of the matrix elements $\langle a| \hat{V}|b\rangle$, for $b \neq a$ are {\em randomized} in the continuous limit, i.e., in the limit where $N$ and $V$ goes to infinity, with $N/V$ constant. By `randomized', we mean that the phases of $\langle a| \hat{V}|b\rangle$ do not exhibit any periodicity, or quasi-periodicity as $b$ varies. Hence, the summation over $b$ is a sum of many random phases, and is expected to be much smaller than the term for $a = b$ that involves no such phases. Eq. (\ref{rpa}) then follows.

  The RPA was initially postulated in systems with  a large but finite number of degrees for freedom, in condensed matter \cite{BohmPi, GeBr} and
in nuclear physics \cite{Bloch, Nami}. However, it  also applies to systems with an infinite number of degrees of freedom, i.e., quantum fields.  Indeed, the name `self-energy'  for the function (\ref{sigmaa}) originates from quantum field theory.

By the RPA,     odd-order terms in the expansion of $G_a(z)$  vanish. Furthermore, the terms of order $2n$ for $n$ integer equal $\Sigma_a(z)^n/(z-E_a)^{n+1}$. Hence, $G_a(z)$ is give by  a geometric series,
\begin{eqnarray}
G_a(z) = \frac{1}{z-E_a} \sum_{n=0}^{\infty} \frac{\Sigma_a(z)^n}{(z-E_a)^n} = \frac{1}{z-E_a} \frac{1}{1 - \frac{\Sigma_a(z)}{z-E_a}} = \frac{1}{z - E_a - \Sigma_a(z)}. \label{asas}
\end{eqnarray}

 Eq. (\ref{asas}) is accurate to order $\lambda^2$. Hence, it can be obtained without the RPA, solely by a perturbative analysis. Most treatments of RPA assume a weaker condition that Eq. (\ref{rpa}), so that the resulting expression for  $G_a(z)$ involves a  self-energy function that coincides with Eq. (\ref{sigmaa}) only to second order in $\lambda$---for details, see,   Ref. \cite{NNP96}. In the present treatment, the RPA gives the same results with a second-order perturbative expansion. In general, it has a larger domain of validity.

\subsection{Structure of the self-energy function}
Eqs. (\ref{asas}) and (\ref{ampll}) imply that
\begin{eqnarray}
{\cal A}_a(t) = \frac{i}{2 \pi  } \lim_{\epsilon \rightarrow 0} \int_{-\infty }^{\infty } \frac{dE e^{-iEt}}{E + i \epsilon- E_a - \Sigma_a(E+i\epsilon)}. \label{ampldec}
\end{eqnarray}
If the self-energy function $\Sigma_a(z)$ were analytic, the integral (\ref{ampldec}) could be evaluated by integrating     along the contour of Fig. 1, and using Cauchy's theorem. However, this is not the case, the self-energy function is discontinuous and it may contain poles or branch points.

To see this, consider the definition     (\ref{sigmaa}) of  $\Sigma_a(z)$. Since the spectrum of $\hat{H}_0$ is continuous for energies larger than $\mu$, $\Sigma_a(z)$ is
     divergent along the half-line $D = \{ z \in {\pmb C}|\; \mbox{Re} z > \mu, \mbox{Im}z = 0\}$. Shifting the Hamiltonian by a constant term, we can always choose    $\mu = 0$, so that $D = {\pmb R}^+$.

To analyze the behavior of  $\Sigma_a(z)$ near $D$, we define
\begin{eqnarray}
\Sigma_a(E^{\pm}) = \lim_{\eta \rightarrow 0} \Sigma_a(E\pm i\eta),
 \end{eqnarray}
for $\eta > 0$. By definition,
\begin{eqnarray}
\mbox{Re} \Sigma_a(E\pm  i \eta) = \sum_b \frac{|V_{ab}|^2(E-E_b)}{(E  - E_b)^2 +  \eta^2}.
\end{eqnarray}
It follows that $\mbox{Re} \Sigma_a(E^+) = \mbox{Re} \Sigma_a(E^-)$, leading to the definition of the {\em level-shift function}

\begin{eqnarray}
F_a(E) := \mbox{Re}  \Sigma_a(E^{\pm})  \label{faE}
\end{eqnarray}
On the other hand, the imaginary part of $\Sigma_a(z)$
\begin{eqnarray}
\mbox{Im} \Sigma_a(E\pm  i \eta) = \mbox{Im} \sum_b \frac{|V_{ab}|^2}{E  - E_b + i \eta} = \mp \eta \sum_b \frac{|V_{ab}|^2}{(E  - E_b)^2 +  \eta^2}, \label{imsa}
\end{eqnarray}
 is discontinuous as the half-line $D$ is crossed.  Eq. (\ref{imsa}) implies that   $\mbox{Im} \Sigma(E^-) = - \mbox{Im} \Sigma(E^+)\geq 0$.

We define the {\em decay function}
\begin{eqnarray}
\Gamma_a(E) := 2 \mbox{Im}  \Sigma(E^-) > 0, \label{gammaE}
\end{eqnarray}
 so that
 \begin{eqnarray}
 \Sigma_a(E^{\pm}) = F_a(E) \mp \frac{i}{2} \Gamma_a(E).
 \end{eqnarray}
  The discontinuity of $\Sigma_a$  across $D$  is
$\Delta \Sigma_a(E) := \Sigma_a(E^+) - \Sigma_a(E^-) = - i \Gamma_a(E)$.
Obviously,  $\Gamma_a(E) = 0$ for $E < 0$.

  Eq. (\ref{ampldec}) becomes
\begin{eqnarray}
{\cal A}_a(t) = \frac{i}{2 \pi  }   \int_{-\infty }^{\infty } \frac{dE e^{-iEt}}{E  - E_a - F_a(E) + \frac{i}{2} \Gamma_a(E)}. \label{ampldecb}
\end{eqnarray}
Since $\Gamma(E) = 0$ for $E < 0$,
\begin{eqnarray}
{\cal A}_a(t) = \frac{i}{2 \pi }  \left( \int_{-\infty }^{0} \frac{dE e^{-iEt}}{E  - E_a - F_a(E)}+ \int_{0 }^{\infty } \frac{dE e^{-iEt}}{E  - E_a - F_a(E) + \frac{i}{2} \Gamma_a(E)} \right). \label{ampldecc}
\end{eqnarray}

On the other hand, Eq. (\ref{propagator2b}) implies that
\begin{eqnarray}
 \lim_{\epsilon \rightarrow 0} \int_{-\infty }^{\infty } \frac{dE e^{-iEt}}{E - i \epsilon- E_a - \Sigma_a(E^-)} = 0,
\end{eqnarray}
hence,
\begin{eqnarray}
  \int_{-\infty }^{0 } \frac{dE e^{-iEt}}{E -   E_a - F_a(E)}  = -  \int_{0 }^{\infty } \frac{dE e^{-iEt}}{E -   E_a - F_a(E) - \frac{i}{2}  \Gamma_a(E)}.
\end{eqnarray}
Substituting into Eq. (\ref{ampldecb}), we obtain
\begin{eqnarray}
{\cal A}_a(t) &=& \frac{i}{2 \pi  }  \int_{0}^{\infty } dE e^{-iEt} \left[ \frac{1}{E  - E_a - F_a(E) + \frac{i}{2} \Gamma_a(E)} - \frac{1}{E  - E_a - F_a(E) - \frac{i}{2} \Gamma_a(E)}\right] \nonumber \\
&=&  \frac{1}{2 \pi }  \int_{0 }^{\infty } \frac{dE \Gamma_a(E) e^{-iEt}}{ [E - E_a - F_a(E)]^2 + \frac{1}{4}[\Gamma_a(E)]^2}. \label{mainampl}
\end{eqnarray}

Eq. (\ref{mainampl}) is the main result of this section, an explicit formula relating the persistence amplitude to the components of the self-energy function.

\subsection{The Wigner-Weisskopf approximation}

The integral  (\ref{mainampl}) involves the functions $F_a(E)$ and $\Gamma_a(E)$ in the denominator. These function are second-order with respect  the perturbation parameter $\lambda << 1$. If  $|F_a(E)| << E_a$ and $\Gamma(E) << E_a$ for $E$ in the vicinity of $E_a$, we can evaluate the persistence amplitude using the {\em Wigner-Weisskopf Approximation} (WWA) \cite{WWA}.

The WWA essentially postulates the substitution of the Lorentzian-like function of $E$ in Eq. (\ref{mainampl}) with an actual Lorentzian. The justification is the following.  The integral (\ref{mainampl}) is dominated by values of $E$  within distance of order $\lambda^2$ from  $E \simeq E_a$. For these values, the integrand is of order $\lambda^{-2}$, otherwise it is of order $\lambda^0$.   Hence, with an error of order $\lambda^2$,
 we   can substitute the energy-shift function $F_a(E)$ with the constant
\begin{eqnarray}
\delta E := F_a(E_a)
\end{eqnarray}
 and the decay function $\Gamma_a(E)$ with the constant
 \begin{eqnarray}
 \Gamma  := \Gamma_a(E_a).
 \end{eqnarray}
 Within an error of the same order of magnitude,
 we extend the range of integration to $(-\infty, \infty)$. Thus, we  obtain an elementary integral
\begin{eqnarray}
{\cal A}_a(t) =  \frac{\Gamma_a}{2 \pi }  \int_{-\infty}^{\infty } \frac{dE   e^{-iEt}}{ (E - E_a -  \delta E )^2 + \frac{1}{4}\Gamma^2} =  e^{ -i (E_a +\delta E ) t - \frac{\Gamma }{2}t},  \label{ampl3}
\end{eqnarray}

 Substituting Eq.  (\ref{ampl3}) in Eq. (\ref{decprob2}), we conclude
\begin{eqnarray}
p(t) = \Gamma  e^{-\Gamma t}.
\end{eqnarray}

 Hence, the WWA leads to an exponential decay law with a decay constant $\Gamma $ that is determined by the imaginary part of the self-energy function. The real part of the self-energy function leads to a shift
 $\delta E $ of the energy level $E_a$, usually referred to as the {\em Lamb shift}.

 The same expression for the decay constant $\Gamma $ is  given by {\em Fermi's golden rule}. To see this, we use
Eq. (\ref{imsa}),
\begin{eqnarray}
\Gamma  =  \lim_{\eta \rightarrow 0} \sum_b \frac{2\eta |V_{ab}|^2}{(E_a  - E_b)^2 +  \eta^2}.
\end{eqnarray}
Since $\lim_{\eta \rightarrow 0} \frac{\eta}{x^2 +\eta^2} = \pi \delta(x)$, we obtain Fermi's decay rate
\begin{eqnarray}
\Gamma  = 2 \pi \sum_{b, E_b = E_a} |V_{ab}|^2. \label{fgg}
\end{eqnarray}

A more rigorous derivation of Eq. (\ref{ampl3}) from Eq. (\ref{ampldecb}) employs the notion of the  {\em van Hove limit} \cite{vanHo}.  This limit is obtained as follows. First, we change the time variable to $\tilde{t} = \lambda^2 t$, and we define $x = (E-E_a)/\lambda^2$. Then, Eq. (\ref{ampldecb}) becomes
\begin{eqnarray}
e^{iE_at} {\cal A}_a(t) =  \frac{i}{2 \pi  } \lim_{\bar{\epsilon} \rightarrow 0} \int_{-\infty }^{\infty } \frac{dx e^{-ix \tilde{t}}}{x + i \bar{\epsilon}-  \tilde{F}_a(E_a+\lambda^2x ) + \frac{i}{2} \tilde{\Gamma}_a(E_a+\lambda^2x)}, \label{aaavH}
\end{eqnarray}
where $\tilde{\Gamma} (E) = \lambda^{-2}\Gamma(E)$ and $\tilde{F} (E) = \lambda^{-2}F (E)$ are of order $\lambda^0$; $\bar{\epsilon} = \epsilon/\lambda^2$ can still be chosen arbitrarily small. The van-Hove limit consists of taking the limit $\lambda \rightarrow 0$ in the r.h.s. of Eq. (\ref{aaavH}), while keeping $\tilde{t}$ constant. Then,
\begin{eqnarray}
e^{iE_at} {\cal A}_a(t) =  \frac{i}{2 \pi  }  \int_{-\infty }^{\infty } \frac{dx e^{-ix \tilde{t}}}{x   - \tilde{F}_a(E_a) + \frac{i}{2} \tilde{\Gamma}_a(E_a)}, \label{aaavH2}
\end{eqnarray}
Eq. (\ref{aaavH2}) can be straightforwardly evaluated using the contour integral of Fig. 1. It leads to  Eq. (\ref{ampl3}). Hence, the WWA  is equivalent to the imposition of the van Hove limit on the decay amplitude.

While the van Hove limit of the persistence amplitude is always well defined, we have to keep in mind that in physical systems $\lambda^2$ is always finite and non-zero. In specific systems, the van Hove limit may misrepresent the form of the persistence amplitude. This is the case if $\Gamma_a(E)$ strongly varies with $E$ within a distance of order $\lambda^2$ from $E_a$.  Hence, taking the van Hove limit is justified only if the self-energy function is sufficiently `flat' in the vicinity of $E_a$ \cite{BaRa}.

\subsection{Beyond   exponential decay }
We derived  exponential decay as a consequence of two approximations, the RPA and WWA. Since the RPA is redundant for a second-order approximation to the self-energy function,  WWA is the only approximation that needs to be considered in the weak coupling regime.

\medskip

\noindent {\em Very early times.} First, we note that exponential decay cannot be valid at very early times. This is a general statement that originates from the definition (\ref{decprob2}).
We Taylor-expand the persistence amplitude around $t = 0$, to obtain    ${\cal A}_{\psi} = 1 - i t \langle \hat{H} \rangle - \frac{t^2}{2} \langle \hat{H}^2 \rangle + \ldots$. Keeping terms up to order    $t^2$,
the probability density becomes
\begin{eqnarray}
|{\cal A}_{\psi} |^2 = 1 - (\Delta H)^2 t^2 + \ldots.
\end{eqnarray}
Eq.  (\ref{decprob2}) implies that
\begin{eqnarray}
p(t) =  (\Delta H)^2 t.
\end{eqnarray}
 It follows that $p(0) =0$, while in   exponential decays,  $p(0) = \Gamma$.  This violation of exponential decay at early times is a special case of the  so called {\em quantum Zeno effect} \cite{MiSu, FP08}, and it has been verified experimentally  \cite{Zeno}.

The early time behavior of a decaying system can also be identified by an uncertainty relation for the persistence probability, first derived by Mandelstam and Tamm \cite{MaTa45}, see, also \cite{Bhatta, Dodo}. For a system in a state $|\psi\rangle$ and Hamiltonian $\hat{H}$, the Kennard-Robertson uncertainty relation gives,
 \begin{eqnarray}
 \Delta H \Delta A \geq \frac{1}{2} |\langle \psi| [\hat{H}, \hat{A}]|\psi\rangle|, \label{kerob}
 \end{eqnarray}
 for any observable $\hat{A}$. For the Heisenberg evolved observable $\hat{A}(t) = e^{i\hat{H}t} \hat{A} e^{-i\hat{H}t}$,  Eq. (\ref{kerob}) implies the so called Mandelstam-Tamm inequality
\begin{eqnarray}
\Delta H \Delta A(t) \geq \frac{1}{2} \large|\langle \psi| \frac{\partial \hat{A}(t)}{\partial t} |\psi\rangle \large| \label{mata1}
\end{eqnarray}
Choosing   $\hat{A} = |\psi\rangle \langle \psi|$, the expectation $ \langle \psi|\hat{A}(t)|\psi\rangle$ coincides with the survival probability of $|\psi\rangle$, which we will denote by $a(t)$.  Hence, Eq.
 (\ref{mata1}) becomes
  \begin{eqnarray}
  \frac{|\dot{a}|}{ \sqrt{a-a^2}} \leq 2\Delta H. \label{mata2}
  \end{eqnarray}

  Eq. (\ref{mata2}) holds for all quantum states and Hamiltonians. It is easy to verify that it is  not compatible with the exponential decay law at early times. Setting $a(t) = e^{-\Gamma t}$, we find that Eq. (\ref{mata2} is satisfied only if $t > \Gamma^{-1} \ln \left[ 1 + \frac{\Gamma^2}{4 (\Delta H)^2}\right]$.

  Assuming that $\dot{a} \leq 0$, we can integrate inequality   (\ref{mata2}), to obtain
\begin{eqnarray}
\cos^{-1} \sqrt{a(t)} \leq  \Delta H t  \label{mata3}
\end{eqnarray}
Eq. (\ref{mata3})  is satisfied trivially for
 $t > \frac{\pi}{2\Delta H}$; for $t < \frac{\pi}{2\Delta H}$, it implies that  $a(t) \geq    \cos^2(\Delta H t)$. The half-life $\tau_h$  of the state is defined by the requirement
  $a(\tau_h) = \frac{1}{2}$. Hence, Eq.  (\ref{mata3}) leads to an uncertainty relation between energy spread and half-life
\begin{eqnarray}
\Delta H \tau_h \geq \frac{\pi}{4},
\end{eqnarray}
 that applies even in regimes where exponential decay fails\footnote{For  $\tau$ such that $a(\tau) = 0$, Eq.  (\ref{mata3}) leads to the uncertainty relation $\Delta H \tau \geq \frac{\pi}{4}$. The time $\tau$ is interpreted as the minimum time required for the system to arrive to a state orthogonal to $|\psi\rangle$, and the uncertainty relation is said to define a limit to the `speed of quantum evolution', and, consequently, to the speed of quantum computation. The inequality has been improved \cite{MaLe, LeTo} in order to also incorporate the case of $\Delta H \rightarrow \infty$. It has then been broadly generalized, for example, to open-system dynamics \cite{MLOS} and classical systems \cite{Shan}---for a review, see, Ref. \cite{DeCa}.  }.

\medskip

\noindent {\em The persistence amplitude in terms of a contour integral. } In order to establish the  range of validity of the WWA, we must evaluate the survival amplitude (\ref{ampldec}) without approximations. To this end, we analytically continue  $\Gamma_a(E)$ to the lower imaginary plane, and we define   $\Sigma^-_a(z) := \Sigma_a(z)$ and $ \Sigma^+_a(z) := \Sigma_a(z) - i \Gamma_a(z)$.   The functions $\Sigma^{\pm}_a(z)$ can be viewed as components of a multi-valued complex function: $\Sigma^+$ corresponds to  the first Riemann sheet, and $\Sigma^-$ corresponds to the second Riemann sheet \cite{colth}.

We define a line integral over the contour $C_-$ of Fig. 1.  The contribution from the  circle at infinity vanishes, hence, taking the limit $\epsilon \rightarrow 0$, we obtain

 \begin{eqnarray}
{\cal A}_a(t) = \frac{i}{2 \pi } \oint_{C_-} dz e^{-izt} \left(\frac{1}{z-E_a - \Sigma^+_a(z)} -  \frac{1}{z-E_a - \Sigma^-_a(z)}  \right)+  I_a(t),  \label{amplnew}
\end{eqnarray}
where $I_a(t)$ is the contribution of the negative real axis\footnote{We can use a different integration contour, consisting of the positive real axis, an arc of the circle at infinity, and a half line $N$ that starts from the latter and ends at the origin---for an example, see, Fig. 5. As long as the contour encloses the physically relevant poles near the positive real axis, the analysis remains unchanged. Then, $I_a(t)$ is a line integral along $N$.
The choice   of the negative imaginary axis for $N$ is particularly useful for calculating the long time limit of Eq. (\ref{amplnew}). In fact, we use it implicitly in the derivation of Eq. (\ref{iat0}).}.
\begin{eqnarray}
I_a(t) = \frac{1}{2\pi }\int_0^{\infty} \frac{dx e^{ixt} \Gamma_a(-x)}{[x+  E_a  +F_a(-x)]^2 + \frac{1}{4}\Gamma_a(-x)^2}. \label{remainder}
\end{eqnarray}

 If $\Sigma^{\pm}_a(z)$ are meromorphic functions in the region of the complex plane enclosed by the contour $C_-$, the line integral in Eq. (\ref{amplnew}) is evaluated by finding the poles of the integrand inside $C_-$. To this end, we must  solve the equation
\begin{eqnarray}
z-E_a - \Sigma_a^{\pm}(z) = 0. \label{poleq}
\end{eqnarray}
We will refer to this contribution to the survival amplitude as the {\em pole term}; we will refer to $I_a(t)$ as the {\em remainder term}.

\medskip

\noindent {\em The pole term.}   Unless $E_a$ is very close to a point of divergence of $\Sigma_a^{\pm} $, we expect that $|\Sigma_a(E_a^+)|/ E_a$  is  much smaller than unity. Hence, there  exist a solution to Eq. (\ref{poleq}) within a distance of order $\lambda^2$ from the point $z = E_a$.  Setting $z = E + \lambda^2 x$, we find that $ \lambda ^2 x = \Sigma^{\pm}_a(E_x + \lambda^2x) = \Sigma_a^{\pm} (E_a) + O(\lambda^4)$. Hence, Eq. (\ref{poleq}) is satisfied for

\begin{eqnarray}
z = E_a + F_a(E_a) \mp \frac{i}{2} \Gamma_a(E_a) + O(\lambda^4). \label{BBpole}
\end{eqnarray}
The pole with the plus sign is outside $C_-$, hence, it does not contribute to the contour integral. The pole with the minus sign reproduces the result of the WWA\footnote{Hence, the WWA justifies a common statement of scattering theory, that unstable particles correspond to  poles of the $S$-matrix on the second Riemann sheet \cite{colth, RNewt}. Indeed, the poles of the S-matrix provide a simple way for identifying the basic properties of an unstable state. However, the $S$-matrix formalism  gives a coarse description of a decay process. It  enforces the constancy of transition rates \cite{Weinberg} by averaging over a long time interval, and, for this reason, it cannot discern transient phenomena, including deviation from exponential decay.}. The associated residue is $[1 - F_a'(E_a) + \frac{i}{2}\Gamma'(E_a)]^{-1}$.

Let all other solutions to Eq.   (\ref{poleq}) inside $C_-$ be  $z = \alpha_i$ and $R_i$ the associated residues. Then, by Cauchy's theorem,

\begin{eqnarray}
{\cal A}_a(t)  = \frac{e^{ -i (E_a +\delta E ) t - \frac{\Gamma }{2}t}}{1- F_a'(E_a) + \frac{i}{2}\Gamma_a'(E_a)}  + K_a(t) + I_a(t), \label{accuracy}
 \end{eqnarray}
 where
 $K_a(t) = \sum_i R_i e^{-i\alpha_i t}$. Hence, the WWA is valid if both terms $K_a(t)$ and $I_a(t)$ are negligible.

  If the self-energy function is analytic in the region enclosed by the contour $C_-$, then, typically,  other roots $z = \alpha_i$ to Eq. (\ref{poleq}) are further away from the real axis than the perturbative root (\ref{BBpole}). This means that   $|Im \alpha_i| >> \Gamma$. Hence, $K_a(t)$  vanishes much faster than the WWA term, and can be neglected after an initial transient period.

   If
 $\Sigma^{\pm}_a(z)$ has divergences near or on the real axis, then there may exist other roots, as close to the real axis as the perturbative root  (\ref{BBpole}).  There is no general rule here, and, in principle, a case-by-case study is required.
   Of particular importance is the possibility of {\em resonance}. If $E_a$ is sufficiently close to   a pole of $\Sigma_a(z)$, then the condition $|\Sigma_a(E_a^+)|/ E_a <<1$ will not hold. Then, there is no perturbative pole, the WWA fails, and decays are likely to be non-exponential.

\medskip

 \noindent{\em Late times.} The contribution of the pole term to the persistence amplitude drops exponentially.  The remainder term $I_a(t)$ drops as an inverse power law, hence, it dominates at sufficiently late times \cite{Helund, Nami2}.

To see this, we change  variables to $y = x t$, and write Eq. (\ref{remainder}) as

\begin{eqnarray}
I_a(t) = \frac{1}{2\pi t}\int_0^{\infty} \frac{e^{iy} \Gamma_a(-y/t)}{[y/t + E_a +F_a(-y/t)]^2 + \frac{1}{4}\Gamma_a(-y/t)^2  }, \label{eqyb}
\end{eqnarray}
 As $t \rightarrow \infty $, the denominator becomes $[E_a+F_a(0)]^2 + \frac{1}{4}\Gamma_a(0)^2 = E_a^2 +O(\lambda^2)$.
  Hence,

\begin{eqnarray}
I_a(t) \simeq    \frac{1}{2\pi E_a^2 t} \int_0^\infty dy e^{i y} \Gamma_a(-y/t) =  \frac{1 }{2\pi E_a^2  }  \int_0^\infty dx e^{i xt} \Gamma_a(-x). \label{iat3}
\end{eqnarray}

The long -time limit of $I_a(t)  $ is determined by  the behaviour of $\Gamma_a(z)$ near zero.
  Let $\Gamma_a(z) \simeq A z^{n}$ as $x \rightarrow 0$, for some positive constant $A$ and integer $n$. We evaluate the integral $\int_0^\infty dx e^{i xt} (-x)^n$ for imaginary time $t = i \tau$, with $\tau >0$, to obtain $(-1)^n n!\tau^{1+n}$. We analytically continue back to $t$, to obtain

\begin{eqnarray}
{\cal A}_a(t)=  - \frac{A n! }{2\pi E_a^2 i^{n+1} } \frac{1}{t^{n+1}}. \label{iat0}
\end{eqnarray}

The persistence amplitude  drops as an inverse power of   $t$. Hence, in the long-time limit, decays are characterized by an inverse-power law and not an exponential one. In most systems, the inverse-power behavior appears is time-scales too large to be measurable. Nonetheless, it has been experimentally confirmed in luminescence decays of dissolved organic
material \cite{longtime}, with the exponent depending on the material.

\bigskip

\noindent We summarize the results of our analysis.
\\
\noindent 1. The exponential law always fails at very short and very long times.
\\
\noindent 2. If the self-energy function has no divergences near or on the real axis,  the exponential law is very accurate at all intermediate times.
\\
\noindent 3. If the self-energy function has divergences near or on the real axis,  there may be corrections to exponential decay from other poles.
\\
\noindent 4. Exponential decay likely fails on resonance, i.e., for energies near a divergence point of the self-energy function.

\section{Lee's model}
In this section, we present a general model for decays that can be applied to many different physical situations. This model originates from the work of T.D. Lee \cite{Lee}. We shall consider two versions of the model, one where the decay is accompanied by the emission of a bosonic particle and one where the decay is accompanied by the emission of two fermions. The former describes  spontaneous emission of photons, the latter describes beta decay.
\subsection{Bosonic emission}
\subsubsection{Definitions and properties}
We consider decays of the form $A' \rightarrow A + B$, in which the emitted bosonic particle $B$ is much lighter than the particles $A'$ and $A$. Examples of this type of decay are the following.
\begin{itemize}
\item  $A'$ is the excited state of a nucleus, or of an atom, or of a molecule, $A$ is the corresponding ground state and   $B$ is a photon.
\item  $A'$ is a heavy nucleus that decays to  the nucleus $A$ and an alpha particle.
\item  $A'$ and $A$ are baryons and  $B$ is a light meson.
\end{itemize}
The key idea in Lee's model is to ignore all degrees of freedom pertaining to the motion of the  heavy particles. The heavy particles are then treated as a two-level system (2LS).The ground state $|g\rangle$ corresponds to the particle $A$ and the excited state $|e\rangle$ to the particle   $A'$. The $B$ particle is described by a bosonic Fock space ${\cal F}_B$. We   denote the vacuum of ${\cal F}_B$ by
  $|0\rangle$ and the creation and annihilation operators by $\hat{a}_r$ and $\hat{a}^{\dagger}_r$. The latter satisfy the canonical commutation relations
  \begin{eqnarray}
  [\hat{a}_r, \hat{a}_s] = 0, \hspace{0,3cm}   [\hat{a}^{\dagger}_r, \hat{a}^{\dagger}_s] = 0, \hspace{0,3cm}    [\hat{a}_r, \hat{a}^{\dagger}_s] = \delta_{rs}.
  \end{eqnarray}
  The indices $r, s, \ldots$ denote the possible states of a single $B$ particle, i.e., they label a (generalized) basis on the associated single-particle  Hilbert space.

The Hilbert space of the total system is  ${\pmb C}^2 \otimes {\cal F}_B$. The Hamiltonian
  $\hat{H}_L$ of Lee's model consists of three terms
\begin{eqnarray}
\hat{H}_L = \hat{H}_A + \hat{H}_B + \hat{V} ,
\end{eqnarray}
where
\begin{eqnarray}
\hat{H}_A = \frac{1}{2} \Omega (\hat{1} + \hat{\sigma}_3) , \label{vlee1}
\end{eqnarray}
is the 2LS Hamiltonian, and $\Omega$ stands for the energy difference of the two levels;
\begin{eqnarray}
\hat{H}_B =  \sum_r \omega_r \hat{a}^{\dagger}_r \hat{a}_r, \label{vlee2}
\end{eqnarray}
is the   Hamiltonian for non-interacting particles $B$ particles;
\begin{eqnarray}
\hat{V} = \sum_r \left(g_r \hat{\sigma}_+   \hat{a}_r + g^*_r  \hat{\sigma}_-   \hat{a}_r^{\dagger} \right), \label{vlee3}
\end{eqnarray}
  is the interaction term. It describes the excitation of the 2LS accompanied by the absorption of a B particle, and the decay of the 2LS accompanied by the emission of a $B$ particle.
   The coefficients  $g_r$ depend upon the physical system under consideration.

The initial state for Lee's model is
\begin{eqnarray}
|A'\rangle = |e\rangle \otimes |0\rangle,
\end{eqnarray}
i.e., the 2LS is excited and no $B$-particle is present.

 We evaluate the survival amplitude
  ${\cal A}(t) := \langle A' |e^{-i\hat{H}_L t}|A'\rangle$ using the series (\ref{seriesprop}) for $\hat{H}_0 = \hat{H}_A + \hat{H}_B$.  We find that
\begin{eqnarray}
(z-\hat{H}_0)^{-1}\hat{V}(z-\hat{H}_0)^{-1}|A'\rangle =  \frac{1}{z - \Omega} |g\rangle \otimes \sum_r \frac{g_r}{z- \omega_r} \hat{a}^{\dagger}_r|0\rangle,
\end{eqnarray}
hence,  the  matrix elements
 $\langle \psi|\hat{V}|A'\rangle$ are non-zero only for $|\psi\rangle$
are of the form $|g\rangle \otimes \hat{a}^{\dagger}_r|0\rangle$.  In particular, $\langle A'|\hat{V}|A'\rangle = 0$.

The next term in the perturbative series is
\begin{eqnarray}
(z-\hat{H}_0)^{-1}\hat{V} (z - \hat{H}_0)^{-1} \hat{V}(z - \hat{H}_0)^{-1}|A'\rangle = \frac{\Sigma(z)}{(z- \Omega)^2}  |A'\rangle, \label{leetwo}
\end{eqnarray}
where
\begin{eqnarray}
\Sigma(z) = \sum_r \frac{|g_r|^2}{z-\omega_r}, \label{selee}
\end{eqnarray}
is the self-energy function for the initial state  $|A'\rangle$.

Since the second-order term is proportional to the zero-th order one, the above expressions are reproduced to all orders of perturbation. We obtain
\begin{eqnarray}
[(z - \hat{H}_0)^{-1}\hat{V}]^{2n}(z - \hat{H}_0)^{-1} |A'\rangle = \frac{1}{z- \Omega} \left(\frac{ \Sigma(z)}{z- \Omega}\right)^n|A'\rangle \nonumber
\end{eqnarray}
\begin{eqnarray}
[(z - \hat{H}_0)^{-1}\hat{V}]^{2n+1} (z - \hat{H}_0)^{-1} |A'\rangle = \frac{1}{z- \Omega} \left(\frac{ \Sigma(z)}{z- \Omega}\right)^n|g\rangle \otimes \sum_r \frac{g_r \hat{a}^{\dagger}_r}{z- \omega_r } |0\rangle. \nonumber
\end{eqnarray}
Hence,
\begin{eqnarray}
\frac{1}{z -\hat{H}}|A'\rangle =  \frac{1}{z-\Omega} \sum_{n=0}^{\infty} \frac{\Sigma(z)^n}{(z-\Omega)^n} \left(|A'\rangle + |g\rangle \otimes \sum_r \frac{g_r \hat{a}^{\dagger}_r}{z- \omega_r} |0\rangle\right)\nonumber \\
 = \frac{1}{z - \Omega - \Sigma(z)}\left(|A'\rangle + |g\rangle \otimes \sum_r \frac{g_r \hat{a}^{\dagger}_r}{z- \omega_r} |0\rangle\right). \label{decay555}
\end{eqnarray}

We conclude that
\begin{eqnarray}
G(z)   =  \frac{1}{z - \Omega - \Sigma(z)}. \label {asas2}
\end{eqnarray}

We obtained Eq.  (\ref{asas}) by resumming the full perturbative series (\ref{seriesprop}), without any approximation. This means that
 Lee's model incorporates the RPA in its definition.

\subsubsection{Spontaneous emission from atoms}
We will evaluate the self-energy function in a simple model where the emitting particles have zero spin and zero rest mass, i.e., scalar photons \cite{AnHu}. This model ignores the effects of polarization, but otherwise it describes well the emission of photons by excited atoms. The inclusion of polarization changes $\Sigma(z)$  only by a multiplicative factor that  can be absorbed in a redefinition of the coupling constant.

In this model, the basis $r$ corresponds to photon momenta ${\pmb k}$, $\omega_r$ corresponds to $\omega_{\pmb k } = |{\pmb k}|$, and the summation over $r$ corresponds to integration with   measure $\frac{d^3k}{(2\pi)^3}$. If  the size $a_0$ of the emitting atom is much smaller than the wavelengths of the emitted radiation, we can describe the atom-radiation interaction in the dipole approximation  \cite{QO}. Then, the coupling coefficients are  $g_{\pmb k} = \frac{\lambda }{\sqrt{\omega_{\pmb k}}}e^{i {\pmb k}\cdot{\pmb r}}$, where $\lambda $ is a dimensionless constant and ${\pmb r}$ is the position vector of the atom\footnote{The coupling originates from an interaction Hamiltonian of the form $\lambda[\hat{\sigma}_-\hat{\phi}^{(-)}({\pmb r}) + \hat{\sigma}_+\hat{\phi}^{(+)}({\pmb r})] $, where the scalar field components $\hat{\phi}^{(\pm)}  ({\pmb r}) $ are given by Eqs. (\ref{scalar}, \ref{chik}).}.

 We substitute into Eq. (\ref{selee}) for the self-energy function, to obtain
\begin{eqnarray}
\Sigma(z) = \frac{\lambda^2}{2 \pi^2} \int_0^{\infty} \frac{k dk}{z - k}. \label{seel4}
\end{eqnarray}

The integral in Eq. (\ref{seel4}) diverges as $k \rightarrow \infty$.  However, photon energies much larger than $a_0^{-1}$, where $a_0$ is the size of the atom, are not physically relevant. Hence, we regularize the integral (\ref{seel4}) by introducing a high-frequency cut-off $\Lambda >> \Omega$,

  \begin{eqnarray}
   \Sigma (z) = \frac{\lambda^2}{2 \pi^2}  \int_0^{\Lambda} \frac{k  dk}{z-k} = - \frac{\lambda^2}{2 \pi^2} \left[ \Lambda + z (\ln(\Lambda - z) - \ln (-z)) \right]. \label{seel5}
  \end{eqnarray}

  There are many other ways to regularize the integral (\ref{seel4}), for example, by inserting an exponential cut-off function  $e^{-k/\Lambda}$. The choice of regularization does not affect the form of $\Sigma(z)$ for the physical range of values of $z$, i.e., $|z|<< \Lambda$. However, it introduces an arbitrariness in the behaviour of $\Sigma(z)$ for $z$ of the order of $\Lambda$. In particular,   the apparent branch-point at $z = \Lambda$ in Eq. (\ref{seel5}) is a artefact of the regularization. As far as the physically relevant values of $z$ are  concerned,  there is no error in substituting
   $\ln(\Lambda - z) \simeq \ln \Lambda$ in Eq. (\ref{seel5}). Furthermore, it is convenient to absorb the constant term in $\Sigma(z)$ into a redefinition of the frequency  $\tilde{\Omega} = \Omega -  \frac{\lambda^2\Lambda}{2 \pi^2}$.
   With these modifications, Eq. (\ref{seel5}) becomes

      \begin{eqnarray}
   \Sigma (z) =   \frac{\lambda^2  }{2 \pi^2} \left[ -  (\ln\Lambda) z  + z \ln (-z) \right]. \label{seel6}
  \end{eqnarray}

   The logarithm in Eq. (\ref{seel6}) is defined in the principal branch,  i.e., its argument lies in  $(-\pi, \pi]$. When evaluating $\Sigma(E^-)$, we substitute $z = E - i \eta$, for $\eta > 0$. Hence, $\ln(-z) = \ln E + \ln[-(1-i\eta/E)]$. As $\eta \rightarrow 0$,  $1-i\eta/E \simeq e^{-i\eta/E}$. We have two options for the $-1$ term in the logarithm, we can express it either as $e^{i \pi}$ or as $e^{-i \pi}$. The first choice gives $\ln (e^{i (\pi-\eta/E)})$, hence, the argument lies in the principal branch. The second choice gives $\ln (e^{i (-\pi-\eta/E)})$, and the argument lies outside the principal branch. Only the first choice is acceptable. Hence, $\ln[-(1-i\eta/E)] = i(\pi-\eta/E)$, and  $\lim_{\eta \rightarrow 0} \ln[-(E-i\eta)] = \ln E + i \pi$. It follows that

\begin{eqnarray}
\Sigma(E^-) = - \frac{\lambda^2}{2 \pi^2} \left[  E \ln(\Lambda/E) - i \pi E ) \right].
\end{eqnarray}
By Eqs. (\ref{faE}) and (\ref{gammaE}),
\begin{eqnarray}
F(E) &=& - \frac{\lambda^2}{2 \pi^2}     E \ln(\Lambda/E)   \\
\Gamma(E) &=& \frac{\lambda^2}{ \pi}E.  \label{gammaE2}
\end{eqnarray}

Hence, the decay constant in the WWA is
\begin{eqnarray}
\Gamma = \Gamma(\tilde{\Omega}) = \frac{\lambda^2}{ \pi}\tilde{\Omega}.
\end{eqnarray}
The Lamb shift depends on the cut-off parameter $\Lambda$ and it induces a renormalization of the frequency $\tilde{\Omega}$.

\bigskip

\noindent{\em Validity of exponential decay.}
We examine the domain of validity of the WWA. First, we consider the contribution from other poles. To this end, we look for solutions to equation
 \begin{eqnarray}
 z - \tilde{\Omega} - \frac{\lambda^2}{2 \pi^2} z \ln(1 - \Lambda/z) = 0.  \label{sol3}
 \end{eqnarray}
 Eq. (\ref{sol3}) admits one solution  for $z$ near  $\Lambda$. To see this, we write $z = \Lambda ( 1 +x)$ for $|x| << 1$. We obtain
$x   \simeq  e^{-\frac{2\pi^2}{\lambda^2}} << 1$ to leading order in $\lambda^2$.  This solution is possibly an artefact of the regularization, but in any case it corresponds to oscillations much more rapid than any physically relevant time scale. It averages to zero in any measurement with  temporal resolution of order $\sigma_T >> \Lambda^{-1}$. We conclude that the contribution of other poles to the decay probability is negligible.

Next, we evaluate the remainder term, Eq. (\ref{iat0}). By Eq. (\ref{gammaE2}), $A = \frac{\lambda^2}{\pi}$ and $n = 1$. Hence,
\begin{eqnarray}
I(t) = \frac{\Gamma}{2 \pi \tilde{\Omega}^3 t^2}. \label{asyem}
\end{eqnarray}
For sufficiently large $t$ the remainder term dominates over the exponential term $e^{- \Gamma t/2}$ in the persistence amplitude. The relevant time-scale $\tau$ is found from the solution of the equation $|I(\tau) |= e^{- \Gamma \tau/2}$ at large $\tau$. We set $\Gamma \tau/ 2 =x$ and $\alpha = \frac{1}{8\pi} (\Gamma/\tilde{\Omega})^3$, to obtain an equation for $x$,
\begin{eqnarray}
2 \ln x - x =  \ln \alpha. \label{eqqq}
\end{eqnarray}
 Solutions to Eq. (\ref{eqqq}) for different values of $\Gamma/\tilde{\Omega}$ are given in the following able.

 \bigskip

 \begin{tabular}{|c||c|c|c|c|c|c|}
   \hline
   $\Gamma/\tilde{\Omega}$ & $10^{-2}$ & $10^{-3}$ & $10^{-4}$ & $10^{-5}$ & $10^{-6}$ & $10^{-7}$ \\
   \hline
   x & 23.3 & 30.8 & 38.1 & 45.4 & 52.5 & 59.8 \\
   \hline
  $ e^{-\Gamma\tau}$ & $5\cdot 10^{-21}$ & $2 \cdot 10^{-27}$ & $8 \cdot 10^{-34}$ & $4 \cdot 10^{-40}$  & $2 \cdot 10^{-46}$  &$ 10^{-52}$ \\
   \hline
 \end{tabular}

 \bigskip

Even for values of $\Gamma/\tilde{\Omega}$ as large as $0.01$,    the exponential decay law breaks down at a time where less than $1:10^{20}$ of the initial   atoms remains in the excited state. We conclude that any deviations from exponential decay in photo-emission are negligible outside the quantum Zeno regime.

\subsubsection{Emission of a massive boson}
We adapt the model of Sec. 3.1.2 to describe the emission of a   bosonic particle of mass $\mu$.  The model is identical to that of Sec. 3.1.2, except for the form of the energy function $\omega_{\pmb k}$. We assume non-relativistic energies for the product particle, hence,  $\omega_{\pmb k}= \mu + \frac{{\pmb k}^2}{2 \mu}$. This model is a variation of the one in Refs. \cite{Ghiold, AA} that describes  the decay of a heavy baryon to a lighter one   with the emission of a pion. It allows for an analytic calculation of the persistence amplitude at all times. Hence,  we can witness the transition between the exponential and the power-law regimes without worrying about the validity of specific approximation schemes.

The self-energy function is
\begin{eqnarray}
\Sigma(z) = \frac{\lambda^2}{2 \pi^2} \int_0^{\infty} \frac{k^2 dk}{\left(\mu + \frac{k^2}{2\mu}\right) \left(z - \mu - \frac{k^2}{2\mu}\right)} = -  \frac{\sqrt{2}\lambda^2\mu}{\pi^2}\int_0^{\infty} \frac{dxx^2}{(1+x^2)(x^2 - \zeta)},
\end{eqnarray}
where we set $\zeta = z/\mu - 1$ and $x = k/(\sqrt{2}\mu)$.  The integral over $x$ is evaluated to $\frac{\pi}{2}(1 +\sqrt{-\zeta})^{-1}$. Hence,
\begin{eqnarray}
\Sigma(z) =  -  \frac{\lambda^2\mu}{\sqrt{2}\pi} \frac{1}{1 + \sqrt{ - \frac{z - \mu}{\mu}}}. \label{sigmazint}
\end{eqnarray}

In the non-relativistic regime, the relevant values of $z$ satisfy $|z - \mu| << \mu$, so we can approximate $(1 + \sqrt{ - \frac{z - \mu}{\mu}})^{-1}$ with $1 -  \sqrt{ -\frac{z - \mu}{\mu}}$. We shift the  energy of  the ground state by $\mu$, so that the energy of the  product  particle starts at  $z = 0$ rather than at $z = \mu$. We also absorb the constant $\Sigma(0)$ into $\Omega$. Then, the  energy of the 2LS is $\tilde{\Omega} = \Omega - \mu  -  \frac{\lambda^2\mu}{\sqrt{2}\pi}$, and the self-energy function becomes
\begin{eqnarray}
\Sigma(z) =  -  \frac{\lambda^2}{\pi} \sqrt{-\frac{\mu z}{2}} . \label{sigmazin}
\end{eqnarray}
For $E > 0$, the level-shift function $F(E)$ vanishes because $\Sigma(E^{\pm})$ is purely imaginary. The decay function is
\begin{eqnarray}
\Gamma(E) = \frac{\lambda^2}{\pi} \sqrt{2m E},
\end{eqnarray}
and the decay constant  $\Gamma = \frac{\lambda^2}{\pi} \sqrt{2m \tilde{\Omega}}$.

The persistence amplitude (\ref{mainampl}) becomes
\begin{eqnarray}
{\cal A}(t) = \frac{\Gamma}{2\pi} \int_0^{\infty} \frac{dE e^{-iEt} \sqrt{E/\tilde{\Omega}}}{(E - \tilde{\Omega})^2 + \frac{\Gamma^2E}{4\tilde{\Omega}}},
\end{eqnarray}
where we wrote $\Gamma(E) = \Gamma \sqrt{E/\tilde{\Omega}}$. We change variables to $x = E/\tilde{\Omega}$, to obtain
\begin{eqnarray}
{\cal A}(t) = \frac{\sqrt{2}\gamma}{\pi} \int_0^{\infty} \frac{dx e^{-ix (\tilde{\Omega t})} \sqrt{x}}{(x-1)^2+ 2 \gamma^2 x},
\end{eqnarray}
where $\gamma = \frac{\Gamma}{2\sqrt{2}\tilde{\Omega}} << 1$. The denominator is a binomial with two roots, at $x = x_{\pm}:= 1- \gamma^2 \pm i \gamma\sqrt{2-\gamma^2}$. We use the identity
\begin{eqnarray}
\frac{x_+-x_-}{(x-x_+)(x-x_-)} = \frac{1}{x-x_+} - \frac{1}{x-x_-},
\end{eqnarray}
and change variables to $y = x \tilde{\Omega}t$, to obtain

\begin{eqnarray}
{\cal A}(t) = \frac{R\left(  (1- \gamma^2 + i \gamma\sqrt{2-\gamma^2})\Omega t\right) - R\left(  (1- \gamma^2 - i \gamma\sqrt{2-\gamma^2})\Omega t\right)}{2\pi i\sqrt{1 - \frac{1}{2} \gamma^2}\sqrt{\tilde{\Omega t}}}.
\end{eqnarray}
We defined the function
\begin{eqnarray}
R(a) := \int_0^{\infty} \frac{dy e^{-iy}\sqrt{y}} {y - a}. \label{fai}
\end{eqnarray}
For $\mbox{Re} \; a > 0$, the integral (\ref{fai})  can be expressed analytically in terms of the Fresnel integrals $C(x) = \int_0^x ds \cos(s^2)$ and $S(x) =  \int_0^x ds \cos(s^2)$. We find \cite{ASt},
\begin{eqnarray}
R(a) = (1-i) \sqrt{\frac{\pi }{2}}-\pi  e^{-i a} \sqrt{-a}-(1+i)e^{-i a}\pi \sqrt{a}\left[C\left(\sqrt{\frac{2}{\pi }} \sqrt{a}\right)+i S\left(\sqrt{\frac{2}{\pi }} \sqrt{a}\right)\right]. \label{closedat}
\end{eqnarray}
Eq. (\ref{closedat}) is a closed expression for the persistence amplitude that is valid for all times. The logarithm of the persistence probability $|{\cal A}(t)|^2$ as a function of time is plotted in Fig. 2. The agreement with exponential decay is excellent until $t \simeq 15 \Gamma^{-1}$. The transition to power-law decay is accompanied by increasingly stronger oscillations that originate from the asymptotic behavior of the Fresnel integrals.

It is important to emphasize that the oscillations of Fig. 2 signify the breakdown not only of exponential decay, but of the  persistence probability method. They imply that the probability density (\ref{decprob2}) takes negative values. Hence, the   method does not correlate with the experimental records, namely, the number of product particles recorded at each moment of time. The result strongly suggests that the persistence probability method is not reliable outside the regime of exponential decay.

\begin{figure}
\includegraphics[height=6cm]{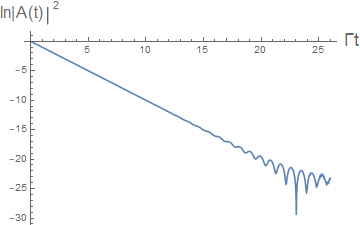}
\caption{The persistence probability as a function of $\Gamma t$ for $\gamma = 0.05$. }
\end{figure}

\subsection{Fermionic emission}
\subsubsection{The Hamiltonian}
Next, we consider decays of the form $A' \rightarrow A + B_1 + B_2$, in which   $B_1$ and $B_2$ are fermionic particles, much lighter than $A'$ and $A$. The most important example of this type is beta decay, where $A'$ and $A$ are nuclei, $B_1$ is an electron (or positron) and $B_2$ is an anti-neutrino (or a neutrino). Again the nucleus is described as a 2LS, with a ground state $|g \rangle$ and an excited state $|e\rangle$.

A fermionic Fock space ${\cal F}_1$ is associated to the  particle $B_1$ and a fermionic Fock space ${\cal F}_2$ is associated to the particle $B_2$. The corresponding ground states are $|0\rangle_1$ and $|0 \rangle_2$, respectively. The creation and annihilation operators on ${\cal F}_1$ will be denoted as $\hat{c}_{r}$ and $\hat{c}_{r}^{\dagger}$, and the creation and annihilation operators  on ${\cal F}_2$ as $\hat{d}_{l}$ and $\hat{d}_{l}^{\dagger}$. They satisfy the canonical anti-commutation relations
\begin{eqnarray}
[\hat{c}_r, \hat{c}_s]_+ = [\hat{c}_r^{\dagger}, \hat{c}^{\dagger}_s]_+ = 0, \hspace{0.4cm} [\hat{c}_r, \hat{c}^{\dagger}_s]_+ = \delta_{rs}
\end{eqnarray}
\begin{eqnarray}
[\hat{d}_l, \hat{d}_m]_+ = [\hat{d}_l^{\dagger}, \hat{d}^{\dagger}_m]_+ = 0, \hspace{0.4cm} [\hat{d}_l, \hat{d}^{\dagger}_m]_+ = \delta_{lm}.
\end{eqnarray}
In the above, $r$ and $s$ are labels of a basis on the Hilbert space of a single $B_1$ particle;  $l$ and $m$   are labels of a basis on the Hilbert space of a single $B_2$ particle. The Hilbert space of the total system is  ${\pmb C}^2 \otimes {\cal F}_1 \otimes {\cal F}_2$

The Hamiltonian for this system again consists of three parts $\hat{H}_L = \hat{H}_A + \hat{H}_B + \hat{V}$, where
\begin{eqnarray}
\hat{H}_A &=& \frac{1}{2} \Omega (\hat{1} + \hat{\sigma}_3)  , \label{vleeb1}\\
\hat{H}_B &=&   \sum_r \omega_r \hat{c}^{\dagger}_r \hat{c}_r   +     \sum_r \tilde{\omega}_l \hat{d}^{\dagger}_l \hat{d}_l  , \label{vleeb2}\\
\hat{V} &=& \sum_{r,l} \left(g_{r,l} \hat{\sigma}_+   \hat{c}_r   \hat{d}_l + g_{r,l}^*  \hat{\sigma}_-   \hat{c}_r^{\dagger}  \hat{d}_l^{\dagger} \right). \label{vleeb3}
\end{eqnarray}
In Eq. (\ref{vleeb2}), $\omega_r$ and $\tilde{\omega}_l$ are energy eigenvalues of a single $B_1$ and $B_2$ particle, respectively. The coefficients $g_{r, l}$ are model-dependent.

The self-energy function  for an initial state $|A'\rangle = |e\rangle \otimes |0\rangle_1 \otimes |0\rangle_2$ is
\begin{eqnarray}
\Sigma(z) = \sum_{r, l} \frac{|g_{r,l}|^2}{z - \omega_r - \tilde{\omega}_l}. \label{sig2f}
\end{eqnarray}

\subsubsection{Beta decay}
We consider a simplified model for beta decay, in which both emitted particles have zero spin and mass. This model is similar to the original Fermi theory of weak interactions \cite{Fermi, Wilson}.
 The zero-mass approximation is reasonable, if the energy $\Omega$ is much larger than the masses of the emitted particles, as is often the case in beta decay. The zero spin approximation is bad; spin is important in the weak interactions.

In this model, the basis $r$ corresponds to momenta ${\pmb p}$, the basis $l$ to momenta ${\pmb q}$, $\omega_r$
  corresponds to $\omega_{\pmb k } = |{\pmb p}|$, and $\tilde{\omega}_l$
  corresponds to $\tilde{\omega}_{\pmb q } = |{\pmb q}|$. The summation over $r$ corresponds to integration over  $\frac{d^3p}{(2\pi)^3}$ and the summation over $s$ corresponds to integration over  $\frac{d^3q}{(2\pi)^3}$. The appropriate constants $g_{r,l}$ for beta decay are
 \begin{eqnarray}
  g_{{\pmb p}, {\pmb q}} = \frac{1}{\mu^2}e^{i({\pmb p}+{\pmb q})\cdot{\pmb r}}
  \end{eqnarray}
   where $\mu$ has dimensions of mass\footnote{In terms of the standard parameters of the theory of weak interactions, $\mu^{-2} = G_F V_{ud} {\cal M}_n$, where $G_F$ is Fermi's constant, $V_{ud}$ is an element of the Kobayashi-Maskawa matrix, and ${\cal M}_n$ the amplitude for the nuclear transition \cite{nuclear}. The dimensionality of $\mu$ comes from the inverse mass squared dimension of Fermi's constant.   } and   ${\pmb r}$ is the position vector of the atom.

   We substitute to  Eq. (\ref{sig2f}), to obtain
  \begin{eqnarray}
  \Sigma(z) = \frac{1}{64 \pi^6 \mu^4} \int \frac{d^3p d^3q}{z - |\pmb p| - |{\pmb q}|} = \frac{1}{16 \pi^4 \mu^4} \int_0^{\infty}p^2dp \int_0^{\infty} q^2 dq  \frac{1}{z-p-q}.
  \end{eqnarray}
We change the integration  variables to $y = p+q, \xi = p-q$. Then,
  \begin{eqnarray}
  \Sigma(z) =   \frac{1}{256 \pi^4 \mu^4} \int_0^{\infty}\frac{dy}{z-y} \int_0^y d\xi (y^2-\xi^2)^2 = \frac{1}{480\pi^4 \mu^4}\int_0^{\infty} \frac{dy y^5}{z-y}, \label{sz2f}
  \end{eqnarray}
The integral in Eq. (\ref{sz2f}) diverges, so we introduce a high-frequency cut-off $\Lambda << \mu$ and restrict the integration over $y$ to the interval $[0, \Lambda]$. Thus, we obtain
\begin{eqnarray}
\Sigma(z) = - \frac{1}{240 \pi^4 \mu^4} \left[ \Lambda^5\left(\frac{1}{5} + \frac{z}{4 \Lambda}  + \frac{z^2}{3\Lambda^2} + \frac{z^3}{2\Lambda^3} + \frac{z^4}{\Lambda^4}\right)+z^5 \ln \Lambda -z^5 \ln(-z)   \right].
\end{eqnarray}
where we approximated  $\ln(\Lambda- z) \simeq \ln \Lambda$.

As in Sec. 3.1.2, the branch point $z = 0$ is logarithmic. Following the same procedure, we evaluate the level-shift and decay functions
 %\begin{eqnarray}
% \Sigma(E^-) = - \frac{\Lambda^5}{1200\pi^4 \mu^4} \left(1+ \frac{5E}{4 \Lambda}  + \frac{5E^2}{3\Lambda^2} + \frac{5E^3}{2\Lambda^3} + \frac{5E^4}{\Lambda^4}\right) -  \frac{E^5}{240 \pi^4 \mu^4}\left( \ln\frac{\Lambda}{E} - i \pi \right).
% \end{eqnarray}

%By Eqs. (\ref{faE}) and (\ref{gammaE}),
\begin{eqnarray}
F(E) &=& - \frac{\Lambda^5}{1200\pi^4 \mu^4} \left(1+ \frac{5E}{4 \Lambda} + \frac{5E^2}{3\Lambda^2} + \frac{5E^3}{2\Lambda^3} + \frac{5E^4}{\Lambda^4}\right) - \frac{E^5}{240 \pi^4 \mu^4}  \ln\frac{\Lambda}{E}
\\
\Gamma(E) &=& \frac{E^5}{120 \pi^3 \mu^4}.  \label{gamma2ef}
\end{eqnarray}
The beta decay rate is
\begin{eqnarray}
\Gamma = \Gamma(\Omega) =  \frac{\Omega^5}{120 \pi^3 \mu^4}
\end{eqnarray}

\section{Resonant decays}
In this section, we consider decays in presence of resonance. To this end, we study  a  variation of the spontaneous emission model of Sec. 3.1.2, in which the 2LS is  within a cavity, consisting of two (infinite) parallel metal plates at distance $L$.  The cavity is perfect, so the field satisfies Dirichlet boundary conditions on the plates.   The model has been studied in Refs. \cite{AnHu, Cum}, but most of the results presented here are new.

\subsection{The self-energy function}

In the directions parallel to the cavity the photon momenta take continuous values. The momentum in the perpendicular direction is an  integer multiple  of the fundamental frequency
\begin{eqnarray}
\omega_0 = \frac{\pi}{L}.
\end{eqnarray}
  Thus, the index $r$ of the bosonic Lee model corresponds to the pair $({\pmb k}, n)$, where ${\pmb k}$ is a two-dimensional vector parallel to the plates and $n = 0, 1, 2, \ldots$. The energies are $\omega_{{\pmb k},n} = \sqrt{{\pmb k}^2 +n^2 \omega_0^2}$ and the mode summation corresponds to $\sum_{n=0}^{\infty} \int \frac{d^2k}{(2\pi)^2}$.

 The coupling constants have the same dependence on energy as in Sec. 3.1.2, but we express them as
  \begin{eqnarray}
  g_{{\pmb k},n} = \lambda \sqrt{\frac{\omega_0}{\omega_{{\pmb k},n}} },
  \end{eqnarray}
  in terms of the dimensionless constant $\lambda$.

Then, the self-energy function takes the form
\begin{eqnarray}
\Sigma(z) = \frac{\lambda^2 \omega_0}{2\pi} \sum_{n=0}^{\infty}\int_0^{\infty} \frac{kdk}{\omega_{{\pmb k},n}(z- \omega_{{\pmb k},n})}  = - \frac{\lambda^2 \omega_0}{2\pi}  \sum_{n=0}^{\infty}  \int_{n \omega_0 -z}^{\infty} \frac{dx}{x}, \label{szcav0}
\end{eqnarray}
 where in the last step we set $x = \omega_{{\pmb k},n} - z$.

 The integral for $\Sigma(z)$ in Eq. (\ref{szcav0}) diverges at high energies. We regularize by introducing a high-energy cut-off $\Lambda$ in the integral over $x$, and a maximum integer $N = \Lambda/\omega_0$ in the summation over $n$.  Then,
 \begin{eqnarray}
 \Sigma(z) &=& -\frac{\lambda^2 \omega_0}{2\pi} \left[ N \ln \frac{\Lambda}{\omega_0} - \sum_{n=0}^{N} \ln(n - z/\omega_0)\right]   \nonumber \\
 &=& -\frac{\lambda^2 \omega_0}{2\pi} \left[\frac{\Lambda}{\omega_0}  \ln \frac{\Lambda}{\omega_0} - \ln\Gamma(N - z/\omega_0) +  \ln\Gamma( -z/\omega_0) \right],
 \end{eqnarray}
 where $\ln\Gamma(z)$ is the logarithmic gamma function. Since the physical values of $z$ are much smaller than $\Lambda$,  we use the asymptotic form of the logarithmic gamma function (Stirling's formula)
 \begin{eqnarray}
   \ln\Gamma(z) \simeq z \ln z -z, \label{stirling}
 \end{eqnarray}
to obtain
\begin{eqnarray}
\Sigma(z) =  - \frac{\lambda^2\Lambda }{2\pi} -    \frac{\lambda^2\log (\Lambda/\omega_0) }{2\pi}z-  \frac{\lambda^2 \omega_0}{2\pi} \ln\Gamma( -z/\omega_0). \label{szcav}
\end{eqnarray}
 As in Sec. 3.1.2, we  incorporate the constant part of $\Sigma(z)$ into a frequency redefinition: $\tilde{\Omega} =  \Omega - \frac{\lambda^2\Lambda }{2\pi}$, so that

\begin{eqnarray}
\Sigma(z) =  - \frac{\lambda^2\log (\Lambda/\omega_0) }{2\pi}z  - \frac{\lambda^2 \omega_0}{2\pi} \ln\Gamma( -z/\omega_0). \label{szcav2}
\end{eqnarray}

The logarithmic gamma function has no poles, but it has infinitely many branch points at all negative integers. The identity $\Gamma(z+1) = z \Gamma(z)$ implies that
\begin{eqnarray}
\ln \Gamma(-x) = - \ln(-x) - \ln(-x+1) - \ldots - \ln(-x + [x]) +\ln\Gamma[-x+[x]+1], \label{lngaid}
\end{eqnarray}
for $x > 0$ and where $[x]$ is the integer part of $x$.

Eq. (\ref{lngaid}) implies that
 $\Sigma(E^-)$   involves the sum of  $[E/\omega_0]+1$ logarithms with negative argument. Each logarithm contributes a term $ \pi$ to the imaginary part of $\Sigma(E^-)$, hence,
\begin{eqnarray}
  \Gamma(E)  =  \lambda^2 \omega_0 ([E/\omega_0]+1).
\end{eqnarray}

 The level-shift function $F(E)$ involves a term $ - \ln(E/\omega_0) -   \ln(E/\omega_0 - 1) - \ln(E/\omega_0 - [E/\omega_0]) + \ln \Gamma( [E/\omega_0] + 1 -  E/\omega_0)$, which can be written as
$ \ln \left( \frac{\Gamma(1+ [E/\omega_0] - E/\omega_0)   \Gamma(E/\omega_0 - [E/\omega_0]) }{ \Gamma(E/\omega_0) }\right)$. Hence,
  \begin{eqnarray}
  F(E) =   - \frac{\lambda^2\log (\Lambda/\omega_0) }{2\pi}z + \frac{\lambda^2\omega_0 }{2\pi} \ln \left( \frac{ \Gamma(E/\omega_0) }{ \Gamma(E/\omega_0 - [E/\omega_0])\Gamma(1+ [E/\omega_0] - E/\omega_0)  }\right).
  \end{eqnarray}
  Both $F(E)$ and $\Gamma(E)$ have finite  discontinuities across the resonances $E = n \omega_0$, for $n = 1, 2, \ldots$.

\medskip

\noindent {\em Large cavity.} For a large cavity ($L \Omega >>1$), $\omega_0$ is very small, so we expand the logarithmic gamma function in Eq. (\ref{szcav2}) using  Stirling's formula. Then,
\begin{eqnarray}
\Sigma(z) =   \frac{\lambda^2  }{2\pi}\left[ - ( \ln \Lambda - 1)z + z \ln(-z) \right] . \label{szcav2b}
\end{eqnarray}
Eq. (\ref{szcav2b}) has the same functional dependence on $z$ with the self-energy function of Eq. (\ref{seel6})   that is obtained in absence of a cavity. If the coupling constant and the cut-off parameters are properly redefined, the two expressions coincide.

\medskip

We evaluate the persistence amplitude using with Eq. (\ref{mainampl}). We define $n_c = [\tilde{\Omega}/\omega_0] $ and $x_0 = \tilde{\Omega}/\omega_0 - n_c $. Then, we express the integral $\int_0^{\infty} dE$   as $\sum_{n=0}^{\infty} \int_{n \omega_0}^{(n+1)\omega_0}dE$, where $n = [E/\omega_0]$.
\begin{eqnarray}
{\cal A}_a(t) = \frac{\lambda^2  }{2 \pi} \sum_{n=0}^{\infty}(n+1)e^{-i n \omega_0 t} \int_0^1 dx \frac{e^{-i(\omega_0t)x}}{[x + n - n_c - x_0 - \frac{\lambda^2 }{2\pi} f_n(x)]^2 + \frac{\lambda^4 }{4}(n+1)^2 } \label{offres}
\end{eqnarray}
 where we substituted $ E = \omega_0(n+x)$ and we wrote
\begin{eqnarray}
f_n(x) =   - \ln(\Lambda/\omega_0) (n +x)  + \ln \frac{ \Gamma(n+x)}{\Gamma(1-x)\Gamma(x)}.
\end{eqnarray}
The function $f_n(x)$ diverges near $x = 0$ and near $x = 1$. Using the expansion,  $\Gamma(x) \simeq \frac{1}{x}  + \ldots$ near $x = 0$, we obtain
\begin{eqnarray}
f_n(x) &=&   \ln x  + \ln (n-1)! -   \ln(\Lambda/\omega_0) n \; \; (n > 0), \label{fn0}
\\ f_n(1-x) &=& \ln x + \ln n! -   \ln(\Lambda/\omega_0) (n+1), \label{fn1}
\end{eqnarray}
for $x << 1$.

 The integrals in Eq. (\ref{offres}) are dominated by  values of $x$ for which the denominator is of order $\lambda^2$.
 Their behavior  depends crucially on whether the system is near resonance or not.

\subsection{ Off resonance}

First, we assume that the atomic frequency is far from the cavity's resonances. Hence, $x_0$ and $1 - x_0$ are much larger than $O(\lambda^2)$. The integrals are dominated by their values near solutions to the equation
\begin{eqnarray}
x + n - n_c - x_0 - \frac{\lambda^2 }{2\pi} f_n(x) = 0, \label{eqoll}
\end{eqnarray}
for $x\in [0, 1]$. For $n = n_c$, Eq. (\ref{eqoll}) admits a perturbative solution at $x = x_0 + \frac{\lambda^2 }{2\pi} f_{n_c}(x_0) + O(\lambda^4)$.

There are   other solutions to Eq. (\ref{eqoll}), near the boundaries  $x = 0$ and $x = 1$ where $f_n(x)$ diverges.
 It is easy to show that these solutions lie at distance  smaller than $e^{-\frac{2 \pi x_0}{\lambda^2}}$ or $e^{-\frac{2 \pi (1-x_0)}{\lambda^2}}$ from the boundaries,  and that their contribution to the integral is negligible.

Therefore, the dominant term to the persistence amplitude corresponds to $n = n_c$. Since the integral is dominated by values near the perturbative solution with a width of order $\lambda^2$, we can extend the range of integration to $(-\infty, \infty)$. But then, we recover the integral (\ref{ampl3}) of the WWA, with

\begin{eqnarray}
\delta E &=&  - \frac{\lambda^2 \omega_0\log (\Lambda/\omega_0) }{2\pi} (n_c  + x_0)  +  \frac{\lambda^2\omega_0 }{2\pi} \ln \left( \frac{\Gamma(n_c + x_0) }{\Gamma(1- x_0 )  \Gamma(x_0)  }\right) \\
\Gamma &=&  \lambda^2 \omega_0 (n_c + 1).
\end{eqnarray}

As expected, the persistence amplitude off-resonance does not differ significantly from the persistence amplitude of an atom outside a cavity.

\subsection{Resonance}
The condition for resonance is that either $x_0$ or $1 - x_0$ is of order $\lambda^2$. In the former case,
$\Omega_R$ is just above the resonance frequency $n_R \omega_0$, for $n_R = n_c$; we define the detuning parameter $\delta = x_0$.
In the latter  case, $\Omega_R$ is just below the resonance frequency $n_R \omega_0$, for $n_R = n_c + 1$; we define the detuning parameter as $\delta =   x_0 - 1$.

Two terms in the series (\ref{offres}) dominate. The first corresponds to $n = n_R$. In this term, the denominator of the integral is of order $\lambda^2$ near $x = 0$. Therefore, we can use the approximation (\ref{fn0}) for $f_n(x)$. We change variables to $y = \frac{\lambda^2}{2\pi}x$ and we extend the limit of integration for $y$ form $\frac{2\pi}{\lambda^2}>> 1$ to $\infty$. Then, this term equals $e^{-in_R\omega_0 t} G_-(\frac{\Gamma_0t}{\pi}, -d, (n_R+1)\pi)$, where

\begin{eqnarray}
G_{\pm}(s, a, b) :=  \frac{b}{\pi}  \int_0^{\infty} \frac{dy e^{- i    s y}}{(y \pm \ln y - a)^2  + b^2 },  \label{gsab}
\end{eqnarray}
\begin{eqnarray}
 d := \ln \frac{2 \pi}{\lambda^2} - \frac{2\pi\delta}{\lambda^2} - \ln (n_R-1)!  +   \ln(\Lambda/\omega_0) n_R,
  \end{eqnarray}
  and $\Gamma_0 =   \lambda^2 \omega_0$.

 The second term corresponds to $n = n_R - 1 $. In this term, the denominator is of order $\lambda^2$ near $x = 1$. Again, we can use the approximation (\ref{fn1}) for $f_n(x)$. We change variables to $y = \frac{2\pi}{\lambda^2}(1-x)$ and take the limit of integration for $y$ to $\infty$. Then, this term equals $e^{-in_R\omega_0 t} G_+^*(\frac{\Gamma_0t}{\pi}, d, \frac{1}{2}n_R\pi)$. Hence,
 \begin{eqnarray}
{\cal A} (t) =  e^{-in_R \omega_0 t} \left[ G_-[\frac{\Gamma_0t}{\pi}, -d,  (n_R+1)\pi ] + G^*_+[\frac{\Gamma_0t}{\pi}, d,  n_R \pi ]   \right], \label{atres}
\end{eqnarray}

For general values of $a$ and $b$ the functions $G_{\pm}(s, a,b)$ can only evaluated numerically. In general, they decay with increasing $s$, with some oscillations for positive $a$.  Plots of $G_{\pm}(s, a, b)$ as a function of $s$ for different values of $a$ and $b$ are given in Fig. 3.

\begin{figure}
\includegraphics[height=10cm]{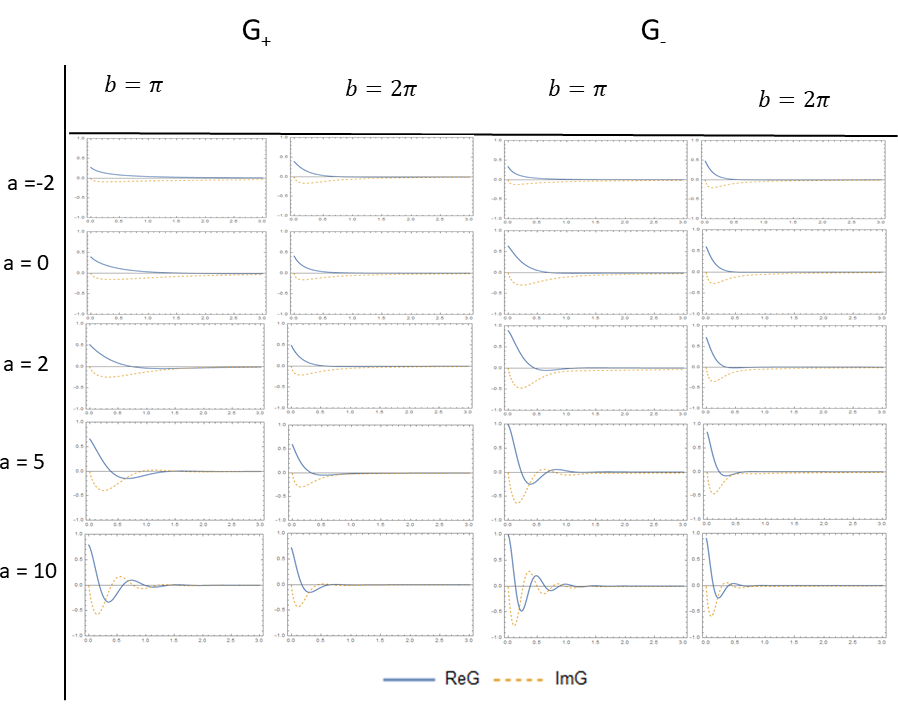}
\caption{The real and the imaginary part of   $G_{\pm}(s, a, b)$ are plotted as  functions of $s$ for different choices of $a$ and   $b$. }
\end{figure}
For  $a >>b $  , the contribution of the logarithm to the integral (\ref{gsab}) is negligible, and the integral is little affected if the range is extended to $(-\infty, \infty)$. Then,
\begin{eqnarray}
G_{\pm}(s, a, b) = \frac{b}{\pi}   \int_{-\infty}^{\infty} \frac{dy e^{-i    s y}}{(y- a )^2  + b^2 } = e^{-b s-ias}
\end{eqnarray}
In contrast, if $|a|>> b$ with $a <0$, $G_{\pm}(s, a, b)$ is of order $(|a|/b)^2$. We readily  verify  that Eq. (\ref{atres}) recovers the exponential decay form for $ |d| >> \pi n_R$.

In all other regimes, the decays are non exponential. This can be seen in  Fig. 4, where the logarithm of the persistence probability   is plotted as a function of $\Gamma_0t/\pi$ for different values of $d$. The graph becomes a straight line for larger values of $d$, signaling exponential decay. Deviations  appear when a tiny fraction of the initial  2LS remains excited. For several values of $d$, the persistence probability is not a decreasing function of $t$. Again, this signifies a failure of the definition (\ref{decprob2}) for the decay probability.

\begin{figure}
\includegraphics[height=7cm]{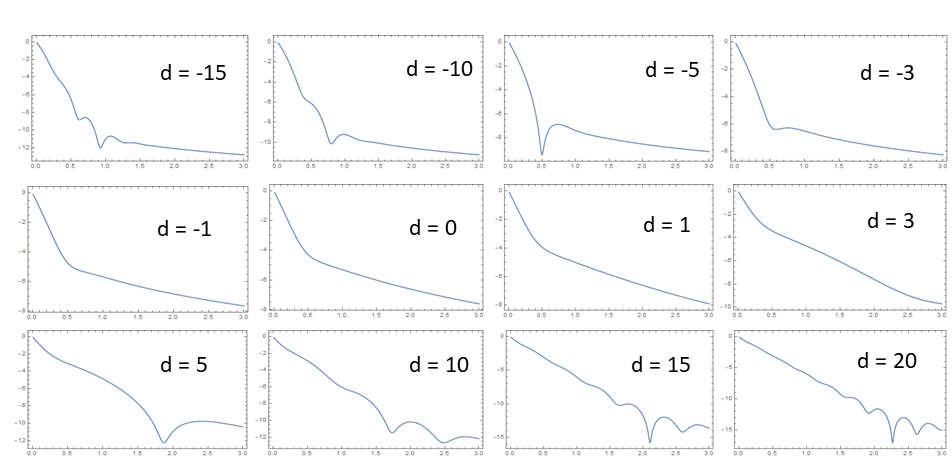}
\caption{The logarithm of the persistence probability $\ln |{\cal A}(t)|^2$ as a function of $\Gamma_0t/\pi$ for different values of $d$ and for $n_R = 1$.   }
\end{figure}

The behavior of the persistence probability derived here is typical  for decaying systems with energies close to a branch point of the self-energy function. 
 In quantum field theory, such branch points appear at energy thresholds, i.e., for energies near  the activation energy $E_0$ of a new decay channel. For example, if the energy of a photon becomes $2m_e$, where $m_e$ is the electrons's mass, the photon decay to a electron-positron pair is possible. In such cases, the persistence amplitude receives two distinct contributions, one from energies slightly beneath and one from energies slightly above the threshold.  For discussions of non-exponential decays due to threshold effects, see, Refs. \cite{LZMM, RZ93, JMSST, DJN09}.

\section{Decay through barrier tunneling}
The methodology developed in Sec. 2 applies to decays that originate from a small perturbation in the Hamiltonian. In this section, we consider non-perturbative decays that can be understood in terms of tunneling. Examples of such decays are the alpha emission of nuclei, tunneling ionization of atoms due to an external field, and leakage of particles from a trap.
We will consider a simple model of a particle in one dimension that mimics the classic treatment of alpha decay by Gamow and by Gurney and Condon \cite{Gamow, GuCo}.

 \subsection{Set-up}

\noindent {\em Dynamics.} We consider a particle in the half-line ${\pmb R}^+ = [0, \infty)$ in presence of  a potential $V(x)$. The potential vanishes outside $ [a, b]$, where $a$ and $b$ are microscopic lengths.

It is convenient to express the potential in terms of the transmission and reflection coefficients  of the  Schr\"odinger operator  $\hat{H} = \frac{\hat{p}^2}{2m}+ V(\hat{x})$ over the full real line.  $\hat{H}$    has two generalized eigenstates $f_{k\pm}(x)$ for each value of energy $E = \frac{k^2}{2m}$.
\begin{eqnarray}
f_{k+}(x) &=& \left\{ \begin{array}{cc} \frac{1}{\sqrt{2\pi}} (e^{ikx} + R_k e^{-ikx}), & x < a\\
 \frac{1}{\sqrt{2\pi}} T_k e^{ikx}& x > b
 \end{array}\right. \nonumber \\
 f_{k-}(x) &=& \left\{ \begin{array}{cc} \frac{1}{\sqrt{2\pi}}  T_k e^{-ikx}, & x < a\\
\frac{1}{\sqrt{2\pi}} (e^{ikx} + \tilde{R}_k e^{ikx})& x > b
 \end{array}\right. \label{f+-}
\end{eqnarray}
where   the complex amplitudes $T_k$, $R_k$ and $\tilde{R}_k$ satisfy
\begin{eqnarray}
|T_k|^2+|R_k|^2 = 1, \hspace{0.75cm}  |R_k| = |\tilde{R}_k|,
  \hspace{0.75cm}  T^*_k R_k +T_k \tilde{R}_k^* = 0. \label{idtyscat}
\end{eqnarray}
$T_k$ is the transmission amplitude, $R_k$ is the reflection amplitude for a right-moving particle and $\tilde{R}_k$ is the reflection amplitude for a left-moving particle.

When the range of $x$ is restricted into the half-line, the generalized eigenfunction $g_k$ of the Schr\"odinger operator   is the linear combination of $f_{k+}$ and $f_{k-}$ that satisfies  $g_E(x) = 0$, i.e.,
\begin{eqnarray}
g_k =  \left( -\frac{T_k}{1+R_k} f_{k+} + f_{k-}\right).
\end{eqnarray}
This implies that
\begin{eqnarray}
g_{k}(x) = \left\{ \begin{array}{cc} -\frac{2i}{\sqrt{2\pi}} \frac{T_k}{1+R_k} \sin kx, & x < a\\
 \frac{1}{\sqrt{2\pi}} \left[ e^{- ikx} - e^{iS_k} e^{ikx} \right] & x > b
 \end{array}\right. , \label{gkx}
\end{eqnarray}
where
\begin{eqnarray}
e^{i S_k} = \frac{T_k^2}{1+R_k} - \tilde{R}_k =  \frac{1+R^*_k}{1+R_k} \left(\frac{T_k^2}{|T_k|^2}\right), \label{eithe}
\end{eqnarray}
is the reflection amplitude of a left-moving particle. The absolute value of the reflection amplitude is unity, because there is no possibility of transmission to $x < 0$.

The eigenfunctions $g_k(x)$ are normalized so that $\int_0^{\infty} g_k(x)^* g_{k'}(x) = \delta(k-k')$. We will represent them by kets $|k\rangle_D$.

\medskip

\noindent {\em Initial state.} We consider an initial state $\psi_0(x)$ with the following four properties. First, it vanishes  outside $[0, a]$. Second, it belongs to the Hilbert space of square-integrable harmonic functions on ${\pmb R}^+$ subject to Dirichlet boundary conditions. Hence, it can be expressed as
\begin{eqnarray}
\psi_0(x) = \sqrt{\frac{2}{\pi}} \int_0^{\infty} \sin(kx) \tilde{\psi}_0(k). \label{Dirich}
\end{eqnarray}
The function $\tilde{\psi}_0(k)$ is defined for $k > 0$. Extending to $k <0$ by $\tilde{\psi}_0(-k) = -\tilde{\psi}(k)$, we can write Eq. (\ref{Dirich}) as $\psi_0(x) = - \frac{i}{\sqrt{2\pi}} \int_{-\infty}^{\infty} dk e^{ikx} \tilde{\psi}_0(k)$. By Eq. (\ref{gkx}),
\begin{eqnarray}
{}_D\langle k|\psi_0\rangle = i \frac{T^*_k}{1+R^*_k}   \tilde{\psi}_0(k).
\end{eqnarray}
Third, we assume that $\psi_0(x)$ is real-valued. For example, $\psi_0$ may be an eigenstate of a  Schr\"odinger operator $\hat{H}'$ with a different potential $U(x)$.  The physical interpretation of this condition is that we prepare the system in an eigenstate of $\hat{H}'$ and at time $t = 0$ we change the potential to $V(x)$, for example, by switching on an external electric field. Fourth,  we assume that   $\psi_0$  has a sharp energy distribution with respect to the Hamiltonian $\hat{H}$: the energy spread for $\psi_0$ is much smaller than the mean energy.

%A family of states that satisfies the above conditions are the eigenstates $\phi_n(x)$ of the infinite potential well in $[0, a]$,
%\begin{eqnarray}
%\phi_n(x) = \sqrt{\frac{2}{a}} \sin\frac{n\pi x}{a}, \label{phin}
%\end{eqnarray}
 %for $n = 1, 2, \ldots$.

 \medskip
 Given an initial state $\psi_0$, we find the state at time $t$
 \begin{eqnarray}
 \psi_t(x) = \int_0^{\infty} dk {}_D\langle k|\psi_0\rangle  g_k(x) e^{ - i\frac{k^2}{2m}t}. \label{evolvt}
 \end{eqnarray}
 The  definition  (\ref{decprob2}) of the decay probability in terms of the persistence probability  is not appropriate for this problem, because the initial configuration of the system does not correspond to an one-dimensional subspace, i.e., the particle may remain confined by the potential without remaining in its initial state. A better choice is perhaps to define the decays probability as $p(t) = - \frac{d}{dt} W(t)$, where
$W(t) = \int_0^a dx|\psi_t(x)|^2$ is the  {\em non-escape probability}, i.e., the probability that the particle is found in $[0, a]$ at time $t$---see, Ref. \cite{CMM95} for the relation between persistence and non-escape probability. We note that the definition of the non-escape probability  is   arbitrary:
 we could have equally well  taken the integration range to $[0, b]$.

 The main drawback of both the non-escape and the persistence probability  is that they  lack a clear operational interpretation.
 We do not carry out measurements of particles at  microscopic scales inside the confining potential, rather we measure the number of particles that have tunneled from the barrier, as recorded by a detector  far from the tunneling region. Furthermore, the persistence probability and the non-escape probability are not always  increasing functions of $t$ \cite{Peshk}. Hence, they may lead to negative values for $p(t)$.

Here, we will proceed  by calculating the probability flux far from the potential region, i.e., we evaluate
\begin{eqnarray}
J(x, t) = \frac{1}{m}\mbox{Im} \left[ \psi^*_t(x) \partial_x \psi_t( x)\right],
\end{eqnarray}
for $x >> b$.

 Eqs. (\ref{gkx}) and (\ref{evolvt}) imply that
\begin{eqnarray}
\psi_t(x)  = -\frac{i}{\sqrt{2\pi}} \int _0^{\infty} dk  \frac{T^*_k}{1+R^*_k}  \left[ e^{iS_k +ikx}-  e^{-ikx}\right] e^{-i\frac{k^2}{2m}t}  \tilde{\psi}_0(k). \label{psitxx}
\end{eqnarray}

For $ x>> b$, there is no stationary phase in the integral involving $e^{-ikx -i\frac{k^2}{2m}t}$. Hence, its contribution to $\psi_t(x)$ is much smaller than that of  the integral involving $e^{ikx -i\frac{k^2}{2m}t}$, which has a stationary phase. By Eq.  (\ref{eithe}),
\begin{eqnarray}
\psi_t(x) = -\frac{i}{\sqrt{2\pi}} \int _0^{\infty} dk \frac{T_k}{1+R_k}  e^{ikx -i\frac{k^2}{2m}t } \tilde{\psi}_0(k). \label{psitx}
\end{eqnarray}
  The integral (\ref{psitx}) contains only out-coming waves, hence, there is  no backflow in the probability current. Note that, in general, $J(x, t)$ can take negative values close to the barrier \cite{Winter}, and hence, it is not reliable for the description of near-field experiments, i.e., when a particle detector is placed at a microscopic distance from the tunneling region.

We expand $(1+R_k)^{-1} = \sum_{n=0}^{\infty}(-R_k)^n$, to write Eq. (\ref{psitx}) as
\begin{eqnarray}
\psi_t(x) = -\frac{i}{\sqrt{2\pi}}\sum_{n=0}^{\infty}\int _0^{\infty} dk T_k(-R_k)^n  e^{ikx -i\frac{k^2}{2m}t } \tilde{\psi}_0(k). \label{psitx1b}
\end{eqnarray}

\subsection{Exponential decay }
Eq. (\ref{psitx}) is accurate for the asymptotic behavior of the wave-function at $x >> b$. To proceed further, we exploit the fact that $\psi_0(k)$ is strongly peaked about a specific value $k_0$, and we evaluate Eq. (\ref{psitx}) in the saddle-point approximation. To this end, we write
  $R_k = -|R_k|e^{i\phi_k}$ and $T_k = |T_k|e^{i\chi_k}$, so that Eq. (\ref{psitx1b}) becomes
\begin{eqnarray}
\psi_t(x) =  - \frac{i}{\sqrt{2\pi}}  \sum_{n=0}^{\infty}  \int _0^{\infty}dk |T_kR_k^n|  e^{ikx -i\frac{k^2}{2m}t + i \chi_k + i n \phi_k} \tilde{\psi}_0(k). \label{psitx2}
\end{eqnarray}
Then, we extend the range of integration of $k$ to $(-\infty, \infty)$ setting $\tilde{\psi}_0(-k) =  - \tilde{\psi}_0(k)$ for negative $k$. The integral is not affected, because the additional terms involve a term $e^{-i|k|x -i\frac{k^2}{2m}t}$, with no stationary phase. Then, we  approximate $|T_k| \simeq |T_{k_0}|, |R_k| \simeq |R_{k_0}|$, $k^2 \simeq k_0^2 + 2 k_0 (k-k_0)$, $\phi_k \simeq \phi_{k_0}+ \phi'_{k_0} (k-k_0)$, $\chi_k \simeq \chi_{k_0}+ \chi'_{k_0} (k-k_0)$. The resulting integral is simply the inverse Fourier transform of $\tilde{\psi}_0(k)$. Hence,
\begin{eqnarray}
\psi_t(x) =  T_{k_0}  e^{ik_0x -i\frac{k_0^2}{2m}t } \sum_{n=0}^{\infty} (-R_{k_0})^n \psi_0(x - \frac{k_0t}{m} + \chi'_{k_0} + n \phi'_{k_0}) \label{psitx3}
\end{eqnarray}

Eq. (\ref{psitx3}) has a natural interpretation in terms of classical concepts. The particle makes successive attempts to cross the barrier at $x = a$. On failure, it is reflected back,  it is reflected again at $x = 0$, and then it makes a new attempt. The $n$-th term in the sum of Eq. (\ref{psitx3}) is the amplitude associated to a particle that succeeded in crossing the barrier at its
 $(n+1)$-th attempt: it is proportional to $T_{p_0}$ (one success) and to $R^n_{p_0}$ (after $n$ failures).

Since $\psi_0(x)$ has support only on $[0, a]$, $\psi_t(x)$ vanishes for $t < t_0 := m (x -a + \chi_{k_0}')/p_0$. The time-scale $t_0$ has an obvious classical interpretation: it is the time it takes a particle inside the barrier region to traverse the distance to point $x$. The term $\chi_{k_0}$ corresponds to the Wigner-Bohm time delay due to the particle crossing the
classically forbidden region \cite{BohmWig}. We rewrite Eq. (\ref{psitx3}) as
\begin{eqnarray}
\psi_t(x) =  T_{k_0}  e^{ik_0x -i\frac{k_0^2}{2m}t } \theta(t-t_0)\sum_{n=0}^{\infty} (-R_{k_0})^n \psi_0\left(a - \frac{k_0(t - t_0 )}{m} + n\Delta x \right)  \label{psitx4}
\end{eqnarray}
where we defined $\Delta x = \phi'_{k_0}$ the position-shift between successive terms in the series (\ref{psitx4}). If $|\Delta x| > a$, the partial amplitudes at different $n$ do not overlap. Hence, there is no quantum interference between different attempts of the particle to cross the barrier. For $|\Delta x| < a$, there is  quantum interference between $M = [a/|\Delta x|]$ successive attempts to cross the barrier.

Since we assumed the initial state $\psi_0$ to be almost monochromatic at energy $\frac{k_0^2}{2m}$, the dominant contribution to the current is $ \frac{k_0}{m} |\psi_t(x)|^2$. Hence,
\begin{eqnarray}
J(t,x) &=& \frac{k_0}{m} |T_{k_0}|^2  \theta(t-t_0) \sum_{n=0}^{\infty} \sum_{\ell=0}^{\infty}(-R_{k_0})^n(-R_{ k_0}^{*})^{\ell}\nonumber \\
&\times& \psi_0(a- \frac{k_0(t - t_0 )}{m} + \ell \Delta x )\psi_0(a- \frac{k_0(t - t_0 )}{m} + n \Delta x ) \label{jxt1}
\end{eqnarray}
Terms in the summation with $|n- \ell| > M$ vanish because the corresponding wave functions do not overlap. Then, we write
\begin{eqnarray}
J(t,x) = \frac{k_0}{m} |T_{k_0}|^2  \theta(t-t_0) \sum_{N=0}^{\infty}|R_{k_0}|^N \rho\left(\frac{k_0(t - t_0 )}{m} - \frac{1}{2} N\Delta x  \right), \label{jtx3}
\end{eqnarray}
where
\begin{eqnarray}
\rho(x) = \sum_{j=-M}^M \psi_0(a - x - \frac{j}{2} \Delta x) \psi_0(a - x +\frac{j}{2} \Delta x) e^{i j\phi_{k_0}}.
\end{eqnarray}
The function $\rho(x)$ is localized within a width of order $a$.

In order to connect Eq. (\ref{jtx3}) with experiments, we have to treat both $x$ and $t$ as macroscopic variables. This means that they can be measured with an accuracy of order $\sigma_X$ and $\sigma_T$, respectively, that is much larger than the microscopic scales that characterize the system. Hence, $\sigma_X >> a$ and $\sigma_T >> ma/k_0$. At such scales, the width of $\rho(x)$ is negligible, and we can substitute it with a delta function,
\begin{eqnarray}
\rho(x) \simeq \alpha \delta(x), \label{Markovtunel}
\end{eqnarray}
where $\alpha = \int_{-\infty}^{\infty}dx \rho(x)$ is a number close to unity. In particular,  $\alpha =1 $ for $M = 0$. In this regime, we can also approximate the sum over $N$ with an integral, so that
\begin{eqnarray}
J(t,x) &=& \alpha \frac{k_0}{m} |T_{k_0}|^2  \theta(t-t_0) \int_0^{\infty}dN |R_{k_0}|^N \delta\left(\frac{k_0(t - t_0 )}{m} - \frac{1}{2} N\Delta x  \right)\nonumber \\  &=& \alpha \frac{|T_{k_0}|^2}{\Delta t}  \theta(t-t_0) e^{\log |R_{p_0}|^2 \frac{(t-t_0)}{\Delta t}} \label{jx5}
\end{eqnarray}
where  $\Delta t = m \Delta x/k_0$ has the classical interpretation as the time between two successive attempts of the particle to cross the barrier. For $|T_{k_0}| << 1$, $\log|R_{p_0}|^2 \simeq -|T_{p_0}|^2$. Then, Eq. (\ref{jx5}) describes exponential decay,

\begin{eqnarray}
J(t,x) =  \alpha \Gamma  e^{-\Gamma  (t-t_0) } \theta (t-t_0), \label{expdec}
\end{eqnarray}
 with a decay constant
\begin{eqnarray}
\Gamma = \frac{ |T_{k_0}|^2}{\Delta t}. \label{Gammat}
\end{eqnarray}
that does not depend on the detailed properties of the initial state.

Note that  exponential decay fails at early times; the derivation of  Eq. (\ref{expdec}) requires that $|t - t_0 |>> \Delta t$.

In deriving the exponential decay law, we employed the saddle point approximation. This it is reasonably accurate for   $\cap$-shaped potentials. In general, it does not
apply  to potentials with multiple transmission and reflection points, like the double well potential \cite{Matsu}. Such potentials may trap the particle in an intermediate region, and they require an analysis of the escape from this region. The  escape satisfies an exponential decay law, except for energies near resonance \cite{AnSav13}.

The exponential decay law also fails at very long times, when wave-function dispersion becomes important. To see this, we change variables to  $y = \frac{k^2t}{2m}$ in Eq. (\ref{psitx}) for $\psi_t(x)$. The dominant term at $t \rightarrow \infty$ is
\begin{eqnarray}
\psi_t(x) =  - i \sqrt{\frac{m}{4\pi t}} A_0  \int_0^{\infty} \frac{dy}{y}  e^{-iy} \tilde{\psi}_0(\sqrt{2my/t}). \label{asymptunel}
\end{eqnarray}
where $A_0 $ stands for $\lim_{k\rightarrow \infty} T_k/(1+R_k)$. In general, $A_0 \neq 0$, as can be readily checked in elementary systems. Hence, the asymptotic behavior of $\psi_t(x)$ depends on  the infrared behavior of $\tilde{\psi}_0$. For  a power law dependence, $\psi_0 \sim k^n$ near $k = 0$, Eq. (\ref{asymptunel}) gives $\psi_t(x) \sim t^{-\frac{n+1}{2}}$. It follows that $J(t, x) \sim t^{-(n+1)}$, i.e., the flux decays  with an inverse power law. We note here that the asymptotic regime captures some of the information of the initial state \cite{MDCG}.

The key property in deriving the exponential decay law is the lack of interference between  different attempts of the particle to cross the barrier. Let us assume that the maximum number of successive attempts that interfere in the probability amplitude is $M$. The decay  time scale $\Gamma^{-1}$ corresponds to $|T_{k_0}|^{-2}$  attempts to cross the barrier. As long as
\begin{eqnarray}
|T_{k_0}|^{-2} >> M, \label{tkm}
\end{eqnarray}
the effects of interference are negligible. This also implies that the particle has very short `memory' about its past attempts to cross the barrier. Hence, the memory time-scale  is much shorter than the decay time-scale. This feature is known as the {\em Markov property}.

In absence of quantum interferences and memory effects,  decays due to tunneling are indistinguishable from classical probabilistic processes that can be described using elementary arguments.  Consider a classical particle that attempts to cross a barrier with probability $w << 1$ of success\footnote{The non-zero probability to cross the barrier needs not be quantum mechanical in origin. The particle may interact with a stochastic environment, such as a thermal bath. Then, the crossing of the  barrier may be due to a random force.}. After $N$ attempts, the survival probability is $(1-w)^N \simeq e^{-Nw}$. If every attempt takes time $\Delta t$, then for $N >> 1$, the system is described by an exponential decay law with constant $\Gamma = w /\Delta t$, in full agreement with Eq. (\ref{Gammat}).

\subsection{Alternative description of tunneling decays}
We can understand the emergence of exponential decay in tunneling using a different argument that does not rely on the saddle-point approximation. Instead, we use the analyticity properties of the time-evolution operator. For the full development of such methods, see, Ref. \cite{RNewt}, and for applications to tunneling decays, see, Refs. \cite{CP75, CMM95, CCM09}.

Assume that we can analytically extend $\tilde{\psi}_0$ to the fourth quadrant of the complex plane. Then, we can write $\psi_t(x) = K(t, x) + I_N(t,x)$, where

\begin{eqnarray}
K(t, x) = \frac{i}{\sqrt{2\pi}}  \oint_C dz  \frac{T_z}{1+R_z}  e^{izx -i\frac{z^2}{2m}t } \tilde{\psi}_0(z), \label{ktn}
\end{eqnarray}
is a line integral along the contour $C$ of Fig. 5, and $I_N(t)$ is the integral across the line segment $N$ of $C$.  By Cauchy's theorem, we can evaluate $K(t,x)$  in terms of the poles of the integrand in the interior of $C$.

\begin{figure}
\includegraphics[height=7cm]{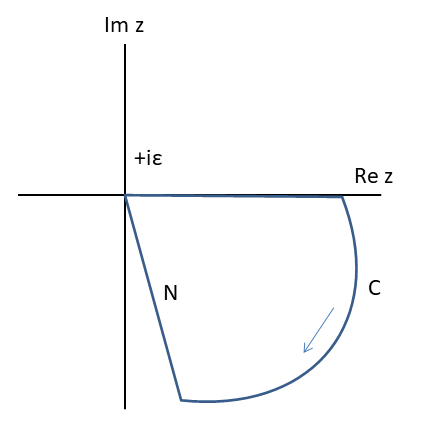}
\caption{The integration contour of the line-integral (\ref{ktn}). The contour is traversed clockwise. }
\end{figure}
Let us denote by $z_n = q_n -i \gamma_n$ the poles of $\frac{T_z}{1+R_z} $ in  the interior of $C$. The integer $n$ labels the poles, in the interior of $C$, $q_n$ and $\gamma_n$ are positive. Then,
\begin{eqnarray}
\psi_t(x) = \sum_n c_n \tilde{\psi}_0(q_n -i \gamma_n) e^{iq_n x - i\frac{q_n^2-\gamma_n^2}{2m}t - \frac{q_n\gamma_n}{m}(t - \frac{mx}{q_n})} + I_N(t),
\end{eqnarray}
for some complex constants $c_n$. Each term in the sum, is suppressed by an exponential factor $\exp[\frac{q_n\gamma_n}{m}(t - \frac{x}{q_n})]$, for $t > \frac{x}{q_n}$. For an almost monochromatic state $\tilde{\psi}_0$ with energy $k_0$, only a small number of poles with $q_n$ near $k_0$ contribute. Furthermore, the contribution   $I_N$ to $\psi_t(x)$ drops exponentially for $t > mx/k_0$, i.e., after the earliest possible time of detection.

For simplicity, let us assume that the contribution of only one pole at $n = n_0$ is significant, and that $q_{n_0} \simeq k_0$. Then, for $t > mx/k_0$,
 \begin{eqnarray}
 \psi_t(x) \sim   e^{ik_0 x - i\frac{k_0^2-\gamma_{n_0}^2}{2m}t - \frac{k_0\gamma_{n_0}}{m}|t - \frac{mx}{k_0}|}.
 \end{eqnarray}
  Therefore, the flux $J(t, x)$ is proportional to $e^{-2 \frac{k_0\gamma_{n_0}}{m} |t - \frac{mx}{k_0}|}$. The decay constant is
  \begin{eqnarray}
  \Gamma = \frac{2k_0\gamma_{n_0}}{m}, \label{decaytb}
  \end{eqnarray}
   and it is determined {\em solely} by the pole of $\frac{T_z}{1+R_z} $ near $z = k_0$. No information about the initial state other than its energy is required.

The above analysis also provides a criterion for the breakdown of exponential decay. If the initial state allows for the contribution of different poles $z_n$, such that there is a significant variation in the values of $\gamma_n$, then corrections to exponential decay, or even its breakdown are possible.

For example, consider an initial state $ \psi_0 = a_1 \psi_1 + a_2 \psi_2$ that is a superposition of  two almost monochromatic states $\psi_1$ and $\psi_2$ with energies $\frac{k_1^2}{2m}$ and $\frac{k_2^2}{2m}$. Furthermore, assume that $k_1$ is close to   one pole of $T_z/(1+R_z)$ at $n = n_1$, and $k_2$ close to  another pole of $T_z/(1+R_z)$ at $n = n_2$, and that there is no overlap. Then, for $t > \max \{ mx/k_1, mx/k_2 \}$,
\begin{eqnarray}
 \psi_t(x) \sim c_1   e^{ik_1 x - i\frac{k_1^2-\gamma_{n_1}^2}{2m}t - \frac{k_0\gamma_{n_1}}{m}|t - \frac{x}{k_1}|} +
  c_2   e^{ik_2 x - i\frac{k_2^2-\gamma_{n_2}^2}{2m}t - \frac{k_0\gamma_{n_2}}{m}|t - \frac{x}{k_2}|},
\end{eqnarray}
for some constants $c_1$ and $c_2$.

The dominant contribution to the current is
\begin{eqnarray}
J(t, x)  &=& |c_1|^2 k_1 e^{- \Gamma_1 |t - \frac{mx}{k_1}|} + |c_2|^2 k_2 e^{- \Gamma_2 |t - \frac{mx}{k_2}|}
\nonumber \\
&+& (k_1 + k_2) e^{- \frac{1}{2} \Gamma_1 |t - \frac{mx}{k_1}| - \frac{1}{2} \Gamma_2 |t - \frac{mx}{k_2}| } \mbox{Re} \left[ c_1c_2^* e^{i \theta(t, x)}\right],
\end{eqnarray}
where   $\Gamma_i = \frac{2k_i\gamma_{n_i}}{m}$, and the interference phase is
\begin{eqnarray}
\theta(t, x) = (k_1 - k_2)x  - \frac{k_1^2 -k_2^2}{2m} t + \frac{\gamma_{n_1}^2 - \gamma_{n_2}^2}{2m}t.
\end{eqnarray}
The flux is characterized by an exponential decay with a periodic modulation due to the energy difference between the interfering states. This is the well-known phenomenon of {\em quantum beats}.

\section{Detection probabilities}
In the previous sections, we employed two methods for constructing the decay probability, namely, persistence probabilities and probabilities currents. Both methods work fine for exponential decays, where the decay probability is determined by a single parameter $\Gamma$. Outside exponential decay they have a restricted domain of validity. The key problem is that they are not guaranteed to define positive-definite probabilities.  This is due to the fact that they do not express probabilities in terms of measurement outcomes for concrete observables, These probabilities are  guaranteed to be positive by the rules of quantum theory.

A rigorous  description of decays requires a consideration of the explicit measurement scheme through which the decay products are detected, and the construction of appropriate measurement observables. The latter correspond to
  positive operators $\hat{\Pi}(t)$, in which the detection time $t$ appears as a random variable. Then, given an initial state $\hat{\rho}_0$, the detection probability $p(t)$ is determined by $Tr\left[\hat{\rho}_0 \hat{\Pi}(t)\right]$.  A scheme for constructing temporal observables of this type has been developed in \cite{AnSav}. Here, we will present an elementary example of such observables that generalizes the well-established photodetection model by Glauber \cite{Glauber}.

Suppose that one of the decay products is a particle that is described by  quantum field  operators $\hat{\phi}({\pmb x})$ and Hamiltonian $\hat{H}$. The field operators are split into a positive frequency part $\hat{\phi}^{(+)}({\pmb x})$ that contains annihilation operators and a negative frequency part $\hat{\phi}^{(-)}({\pmb x})$ that contains creation operators. Consider an elementary apparatus located at a point ${\pmb x}$ that gives a detection signal at time $t$ after having absorbed the incoming particle. The amplitude associated to this process is then proportional to $\hat{\phi}^{(+)}({\pmb x})|\psi_t\rangle$, where  $|\psi_t \rangle$  is the state of the quantum field at time $t$. The probability of detection is the determined by the modulus square of this amplitude. It is given by
  Glauber's formula
\begin{eqnarray}
P(t, {\pmb x})  = C  \langle \psi_t| \hat{\phi}^{(-)}({\pmb x})\hat{\phi}^{(+)}({\pmb x})|\psi_t\rangle, \label{Glauber}
\end{eqnarray}
where   $C$ is a normalization constant.  We do not obtain normalized probabilities, because there is a non-zero probability that the particle will not be detected and this probability depends on the field-state.

Eq. (\ref{Glauber}) was first proposed by Glauber for photodetection. In Glauber's theory,  the role of $\hat{\phi}$ is played by the electric field, and the absorption interaction corresponds to the dipole coupling between the electromagnetic field and a macroscopic detectors.    Glauber's formula is a special case of a larger class of particle detection observables that can be defined in quantum fields \cite{AnSav}.

We will apply Glauber's formula to the bosonic Lee model. The field operators associated to the bosonic creation and annihilation operators are
\begin{eqnarray}
\hat{\phi}^{(+)}({\pmb x}) = \sum_r \hat{a}_r \chi_r({\pmb x}) \hspace{1cm}  \hat{\phi}^{(-)}({\pmb x}) = \sum_r \hat{a}_r \chi^*_r({\pmb x}) \label{scalar}
\end{eqnarray}
where $\chi_r(x)$ are eigenfunctions of the single-particle Hamiltonian. For particles in three dimensions,  $r$ corresponds to the three-momentum ${\pmb k}$, and
\begin{eqnarray}
\chi_{\pmb k}({\pmb x}) = \frac{1}{\sqrt{2 \omega_{\pmb k}} }e^{i{\pmb k} \cdot {\pmb x}}. \label{chik}
\end{eqnarray}

By Eq. (\ref{decay555}),
\begin{eqnarray}
\hat{\phi}^{(+)}({\pmb x}) \frac{1}{z - \hat{H}} |A'\rangle = \frac{V(z; {\pmb x})}{z - \Omega - \Sigma(z)} |g\rangle \otimes  |0\rangle,
\end{eqnarray}
where
\begin{eqnarray}
V_ {\pmb x}(z)  = \sum_r \frac{g_r \chi_r({\pmb x})  }{z- \omega_r}. \label{Vz}
\end{eqnarray}
Then, Eq. (\ref{Glauber}) gives
\begin{eqnarray}
P(t, {\pmb x})  = C  |{\cal B}(t, {\pmb x})|^2, \label{Glauber2}
\end{eqnarray}
where
\begin{eqnarray}
{\cal B}(t, {\pmb x}) =   \lim_{\epsilon \rightarrow 0} \int_{-\infty+i \epsilon }^{\infty+i \epsilon } \frac{dE V_ {\pmb x}(E)  e^{-iEt}}{[E - \Omega - \Sigma_a(E) }.
\end{eqnarray}
Following the same steps that lead to  Eq. (\ref{mainampl}), we find that
\begin{eqnarray}
{\cal B}(t,  {\pmb x}) =
   \int_{0 }^{\infty } \frac{dE e^{-iEt}}{2\pi}  \frac{ \frac{1}{2}\Gamma(E) [V_{\pmb x}^+(E) +V_{\pmb x}^-(E) ] + i [E - \Omega - F(E)][V_{\pmb x}^+(E) - V_{\pmb x}^-(E) ]  }{ [E - \Omega - F(E)]^2 + \frac{1}{4}[\Gamma(E)]^2} \label{mainampl44}
\end{eqnarray}
 where
 \begin{eqnarray}
 V_{\pmb x}^{\pm}(E, {\pmb x}) :=   \lim_{\eta \rightarrow 0 } V_{\pmb x}(E \pm i \eta).
 \end{eqnarray}

We evaluate the amplitude (\ref{mainampl44})  in the WWA, in which the dominant contribution to the integral comes
 from values of $E$ near $\Omega$. Hence,
\begin{eqnarray}
{\cal B}(t,  {\pmb x}) &\simeq&
  \frac{1}{2 \pi }  \int_{-\infty }^{\infty } dE e^{-iEt}  \frac{ \Gamma  [V_{\pmb x}^+(\Omega) +V_{\pmb x}^-(\Omega) ]  + i [E - \Omega - \delta E][V_{\pmb x}^+(\Omega)  - V_{\pmb x}^-(\Omega) ]   }{ [E - \Omega - \delta E]^2 + \frac{1}{4}\Gamma^2} \nonumber \\
&=& \frac{1}{2} e^{-i (\Omega t + \delta E)t - \frac{1}{2}\Gamma t}   V_+(\Omega; {\pmb x})   , \label{mainampl44b}
\end{eqnarray}
where we set $\Gamma = \Gamma(\Omega)$ and $\delta E = F(\Omega)$. Hence, we obtain
\begin{eqnarray}
P(t, {\pmb x})  = \frac{1}{4} C|V_{\pmb x}^+(\Omega) |^2 e^{-\Gamma t}. \label{decayGl}
\end{eqnarray}
The constant $C$ can be determined by normalizing over all particle detection events, i.e., by the requirement that
\begin{eqnarray}
\int_0^{\infty} dt \oint_{S} d^2n  P(t, {\pmb x}) =1  , \label{normalk}
\end{eqnarray}
where $S$ is a two-sphere at distance $r$ from the location of the 2LS, ${\pmb n}$ is a unit vector such that ${\pmb x} = r {\pmb n}$ on $S$.

We conclude  that the WWA guarantees exponential decay with constant $\Gamma$, {\em irrespective of the method used}. However, outside the exponential decay regime the probability density (\ref{Glauber2}) differs significantly from Eq. (\ref{decprob2}). In particular, it is guaranteed to be always positive.

\medskip

As an example, we revisit the photoemission model of Sec. 3.1.2. We employ Eqs. (\ref{chik}) and (\ref{Vz}), to obtain
\begin{eqnarray}
V_{\pmb x}(z) = \frac{\lambda}{8\sqrt{2} \pi^3} \int \frac{d^3k}{|{\pmb k}|(z - |{\pmb k}|)} e^{i {\pmb k} \cdot {\pmb x}}.
\end{eqnarray}
We introduce spherical coordinates $(k, \theta, \phi)$ for ${\pmb k}$, so that
\begin{eqnarray}
V_{\pmb x}(z) = \frac{\lambda}{4\sqrt{2} \pi^2} \int_0^{\infty} \frac{kdk}{z- k} \int_0^{\pi} d\theta \sin \theta e^{ikr \cos \theta} =
 \frac{\lambda}{2\sqrt{2} \pi^2 r} \int_0^{\infty} \frac{dk \sin(kr)}{z-k}, \label{vzr}
\end{eqnarray}
i.e., $V_{\pmb x}(z)$ depends only on the radial coordinate $r = |{\pmb x}|$.

We evaluate the integral (\ref{vzr}) to
\begin{eqnarray}
V_r(z) =  \frac{\lambda}{2\sqrt{2} \pi^2r}  \left[ [\gamma + \ln(-rz) + \mbox{Cin}(rz)] \sin rz - \mbox{si}(rz)\cos rz\right],
 \end{eqnarray}
where the functions
 $\mbox{Cin}$ and $\mbox{si}$ are    defined as
\begin{eqnarray}
\mbox{Cin}(z) = \int_0^z dt \frac{1-\cos t}{t} \hspace{1cm} \mbox{si}(z) = \int_z^{\infty} dt \frac{\sin t}{t}, \label{CiSi}
\end{eqnarray}
and $\gamma$ is the Euler-Mascheroni constant \cite{ASt}.

We straightforwardly evaluate
\begin{eqnarray}
V_r^{\pm}(E;r) = \frac{\lambda}{2\sqrt{2} \pi^2r}  \left[ [\gamma + \ln(rE) + \mbox{Cin}(rE)] \sin rE - \mbox{si}(rE)\cos rE \mp i \pi \sin(rE)\right].
\end{eqnarray}

We assume that the detectors are located at macroscopic distance from the decaying atom, so that $\tilde{\Omega}r >> 1$. Then, the terms involving the trigonometric integrals vanish, and the imaginary part of $V_+(\Omega, r)$  dominates. Eq. (\ref{decayGl}) gives
\begin{eqnarray}
P(t, r) =  \frac{\lambda^2 \sin^2(\tilde{ \Omega} r)}{32 \pi^2 r^2} C e^{-\Gamma t}.
\end{eqnarray}
 The sinusoidal dependence  on $r$ disappears if we average $P(t, r)$ over a thin shell of width $d >> \tilde{\Omega}^{-1}$ at distance $r$, since $\langle\sin^2(\tilde{\Omega}r)\rangle = \frac{1}{2}$. Then, the normalization condition (\ref{normalk}) is satisfied for $C = (16 \tilde{\Omega})^{-1} $.

  Note that outside the exponential decay regime, the probability density (\ref{Glauber2}) leads to different predictions from the persistence probability method. The latter predicts an asymptotic probability density decaying with $t^{-5}$. Eq. (\ref{mainampl44}) leads to an asymptotic decay of ${\cal B}$ with $t^{-2}$, hence,  the probability density (\ref{Glauber2})
decays as $t^{-4}$.

\section{Conclusions}
We presented an overview of the quantum description of decay processes. We showed that the emergence of the exponential decay law is explained in terms of a scale separation. In perturbative decays, the scale separation refers to energy: exponential decays emerge when the released energy associated to the decay  is much larger the energy of the interaction, as described by the self-energy function. In non-perturbative decays, the scale separation refers to time: exponential decay emerges when the decay time-scale is much larger than the time-scale of coherence between different attempts of the particle to cross the barrier.

Exponential decay may be extremely common, but it is not universal. It is not valid at very early and very late times, and in specific systems, it is not relevant at all. There is good experimental evidence for non-exponential decays, some of which pose persistent theoretical puzzles \cite{GSI}. Our increasing access and control of multi-partite/multi-particle systems is expected to uncover further unconventional types of decay---for example,  memory effects due to interaction with an environment \cite{nonMark, BKEC},  effects due to the entanglement of the initial state \cite{AnHu2, CVS17, C18}, or effects from particle statistics in many-particle systems \cite{ADCM, TS11, CL11, DC11, MG11}. Furthermore, studies of    decay dynamics in many-particle quantum systems have demonstrated the need of  a genuinely many-particle characterization \cite{ PSC12, HZHB13, DC16}, i.e., going beyond
description in terms of single-particle observables.  We believe that a significant upgrade of traditional methods for quantum decays will be needed, in order to address such challenges.

\section*{Acknowledgements}
Research was supported by Grant No. E611 from the Research Committee of the University of Patras via the ”K. Karatheodoris” program.

\begin{appendix}
\section{Further results}
The verification of the statements in this Appendix can be used as exercises.

\bigskip

\noindent 1. The persistence amplitude cannot give an exponential decay law at all times \cite{Khalfin}. Let $|a\rangle$ stand for the eigenvectors of the Hamiltonian $\hat{H}$ with eigenvalues $E_a$. Then, the persistence amplitude (\ref{persiss}) can be written as ${\cal A}_{\psi}(t) = \int dE \omega(E)e^{-iEt}$, where
\begin{eqnarray}
\omega(E) = \sum_a \delta(E-E_a) |\langle a|\psi\rangle|^2.
\end{eqnarray}
If the Hamiltonian is bounded from below, with a minimum energy $E_{min}$, then $\omega(E) = 0 $ for $E < E_{min}$. A theorem by Payley and Wiener  \cite{PaWi} asserts that if $\omega(E) = 0 $ for $E < E_{min}$, then its
Fourier transform $A(t)$  satisfies
\begin{eqnarray}
\int_{-\infty}^{\infty} dt \frac{ \log|A(t)|}{1+t^2} < \infty.
\end{eqnarray}
 This implies that $|{\cal A}_{\psi}(t)|$ decays at most  as $e^{-c t^{1 - p}}$ for $c > 0$ and $p >0$, i.e., more slowly than exponential decay.

 \bigskip

\noindent 2.
 A different Hamiltonian defined on the  Hilbert space of Lee's model is the so called spin-boson Hamiltonian, $\hat{H}_{SB}$. The spin-boson model \cite{spbo} is more fundamental, in the sense that it can be derived from first principles with fewer assumptions than Lee's Hamiltonian. The spin-boson Hamiltonian is of the form
  $\hat{H}_{SB} = \hat{H}_A + \hat{H}_B + \hat{V}'$, where $\hat{H}_A$ and $\hat{H}_B$ are given by Eqs. (\ref{vlee1}) and (\ref{vlee2}) respectively, and
\begin{eqnarray}
\hat{V}' =  \hat{\sigma}_1  \sum_a \left(g_r     \hat{a}_r + g_r^*   \hat{a}_r^{\dagger} \right). \label{vsb}
\end{eqnarray}
Since $\hat{\sigma}_1 = \hat{\sigma}_+ + \hat{\sigma}_-$, the interaction term $\hat{V}'$ includes terms  $\hat{\sigma}_+  \hat{a}^{\dagger}_r$ and $\hat{\sigma}_-   \hat{a}_r$, in addition to the ones of  $\hat{V}$  of Εξ. (\ref{vlee3}).

Assuming the RPA, we find that the self-energy function  $\Sigma(z)$ is given by Eq. (\ref{selee}) as is the Lee model. Hence,
\begin{eqnarray}
\langle A' |(z -\hat{H}_{SB})^{-1}|A'\rangle  = \langle A |(z -\hat{H}_L)^{-1}|A'\rangle =   \frac{1}{z - \Omega - \Sigma(z)}.
\end{eqnarray}
 Within the accuracy of the RPA, the spin-boson and Lee's Hamiltonian lead to the same predictions.

 \bigskip

 \noindent 3. The correct electromagnetic description of photonic emission has to take into account photon polarization. In this case, the
basis $r$ corresponds to momenta ${\pmb k}$ and polarization directions $\sigma = 1, 2$; summation over $r$ corresponds to integration over ${\pmb k}$  with   measure $\frac{d^3k}{(2\pi)^3}$ and summation over $\sigma$. The coupling coefficients are the form
\begin{eqnarray}
g_{{\pmb k}, \sigma} = \frac{\lambda }{\sqrt{\omega_{\pmb k}}} [{\pmb n}\cdot{\pmb \omega}_{\sigma} ({\pmb k})]  e^{i {\pmb k}\cdot{\pmb r}}
\end{eqnarray}
 where $\lambda $ is a dimensionless constant, ${\pmb n}$ is a unit vector,  and ${\pmb r}$ is the position vector of the atom. The polarization vectors $\epsilon^i({\pmb k})$ have unit norm, they can be chosen to be real, they are transverse: ${\pmb \epsilon}_{\sigma} ({\pmb k})\cdot {\pmb k} = 0$, and they satisfy
 \begin{eqnarray}
 \sum_{\sigma} \epsilon^i_{\sigma}({\pmb k})  \epsilon^j_{\sigma}({\pmb k})  = \delta^{ij} - \frac{k^ik^j}{{\pmb k}^2}. \nonumber
 \end{eqnarray}
Then, the self-energy function is
   \begin{eqnarray}
   \Sigma (z) = \frac{\lambda^2}{3 \pi^2}  \int_0^{\Lambda} \frac{k  dk}{z-k},
  \end{eqnarray}
 i.e., it differs from Eq. (\ref{seel4}) by a factor $\frac{2}{3}$.

\bigskip

\noindent 4. Consider the model of Sec. 3.1.2 in one spatial dimension, describing, for example, the decay of an atom inside an one dimensional cavity of width much smaller than the wave-length of the emitted radiation. Using the same coupling as in Sec. 3.1.2, the self-energy is
\begin{eqnarray}
\Sigma(z) = \frac{\lambda^2}{\pi} \int_0^{\infty} \frac{dk}{k(z-k)}.
\end{eqnarray}
The integral above is finite at large $k$ but diverges at $k = 0$. W  change the integration range to $[\kappa, \infty)$, where $\kappa$ is an infrared cut-off. Then,
\begin{eqnarray}
\Sigma(z) = -\frac{\lambda^2}{\pi z} \ln \left(1 - \frac{z}{\kappa}\right). \label{sigma1s}
\end{eqnarray}
Eq. (\ref{sigma1s}) implies that $\Gamma(E) = 2\lambda^2/(\kappa + E)$.

\bigskip
\noindent 5. We consider a  two-level atom  of frequency $\Omega$ in an engineered reservoir with a frequency distribution sharply peaked around $\omega_0$. To this end, we employ the model of Sec. 3.1.2 with coupling coefficients $g_{\pmb k} = \frac{\lambda}{\sqrt{\omega_k}} \sqrt{\omega_0 \chi(\omega_{\pmb k}, \omega_0)}$, where $\chi(\omega, \omega_0)$ is a probability distribution on $[0, \infty)$ sharply peaked around $\omega_0$ with a width $\gamma << \omega_0$. The self-energy function equals
\begin{eqnarray}
\Sigma(z) = -\frac{\lambda^2 \omega_0}{2 \pi^2} + \frac{\lambda^2 \omega_0}{2 \pi^2}z \int_{0}^{\infty} \frac{dk f(k, \omega_0)}{z- k}. \label{2LSres}
\end{eqnarray}
Since the integrand in Eq. (\ref{2LSres}) is peaked around $\omega_0$, we can extend the $k$ integration to $(-\infty, \infty)$. In  this approximation, we can choose a Lorentzian function for $\chi$,
\begin{eqnarray}
\chi(\omega, \omega_0) = \frac{1}{\pi} \frac{\gamma}{(\omega-\omega_0)^2 + \gamma^2}.
\end{eqnarray}
Then, 
\begin{eqnarray}
\Sigma(z) = \Gamma \frac{\omega_0- i \gamma}{z - \omega_0 + i \gamma},
\end{eqnarray}
where $\Gamma = \frac{\lambda^2 \omega_0}{2 \pi^2}$.

 $\Sigma(z)$ has no branch points. Hence, the persistence amplitude is solely determined by the solutions of Eq. $z - \Omega- \Sigma(z) = 0$. They correspond to the two roots of the binomial equation $x^2 - (\delta - i \gamma) x - \Gamma (\omega_0 - i \gamma) = 0$, where $x = z - \Omega$,  and
 $\delta := \omega_0 - \tilde{\Omega}$. 
 
 Two limits are particularly interesting. For exact resonance, $\delta = 0$, the roots to leading order in $\gamma/\omega_0$ are $z_{\pm} = \Omega \pm \sqrt{\Gamma \omega_0 - \frac{1}{4}\gamma^2} - i \frac{\gamma}{2}$. The persistence amplitude exhibits oscillations at frequency $\sqrt{4 \Gamma \omega_0 - \gamma^2}$, and decays exponentially with a decay constant $\gamma$. 
 For a sharply monochromatic cavity with $\gamma \rightarrow 0$, both roots are real valued. There is no decay, and the persistence amplitude describes vacuum Rabi oscillations of frequency $\sqrt{\delta^2 + 4 \Gamma^2 \omega_0^2}$.

\bigskip
\noindent 6.    We generalize   Lee's model in order to describe the decay of  an entangled pair of 2LSs. The Hamiltonian   is of the form $\hat{H}_0 + \hat{V}$ with
\begin{eqnarray}
\hat{H}_0= \frac{1}{2} \Omega (\hat{1} + \hat{\sigma}^{(1)}_3) + \frac{1}{2} \Omega (\hat{1} + \hat{\sigma}^{(2)}_3)
 +  \sum_r \omega_r \hat{a}^{\dagger}_r \hat{a}_r,  \label{2h0} \\
\hat{V} = \sum_r \left(g^{(1)}_{r} \hat{\sigma}_+^{(1}   \hat{a}_r + g_{r}^{(1)*}  \hat{\sigma}_-^{(1)}   \hat{a}_r^{\dagger} \right) + \sum_r \left(g_{r}^{(2)} \hat{\sigma}_+^{(2)}   \hat{a}_r + g_r^{*(2)}  \hat{\sigma}_-^{(2)}   \hat{a}_r^{\dagger} \right), \label{2v}
\end{eqnarray}
where the upper indices $(1)$ and $(2)$ label the 2LS.  For identical 2LS the absolute value of the coupling constants $|g^{(i)}_r|$ is the same for both atoms: $|g^{(1)}_r| = |g^{(2)}_r| = g_r$. Hence, we write $g^{(i)}_r = g_r e^{i \Theta^{(i)}_r}$.

We consider an initial state   $|B\rangle \otimes|0\rangle$, where $|0\rangle$ is the field vacuum and $|B\rangle$ is a Bell-type state
\begin{eqnarray}
|B\rangle = \frac{1}{\sqrt{2}}\left(|e,g\rangle \pm |g, e\rangle \right).
\end{eqnarray}

The self-energy function $\Sigma_B$  is
\begin{eqnarray}
\Sigma_B(z) =
\Sigma_0(z) + \frac{1}{2}  [  \Sigma^{(1)}(z) +  \Sigma^{(2)}(z) ], \label{seef2}
\end{eqnarray}
where $\Sigma_0(z)$ is the self-energy function for a single 2LS, given by Eq. (\ref{selee}), and
\begin{eqnarray}
\Sigma^{(i)}(z) = \sum_r \frac{g_r^2e^{2i\Theta_r^{(i)}}}{z- \omega_r}.
\end{eqnarray}

For the  scalar photons  of Sec. 3.1.3,  we substitute $r$  by the continuous momentum variable ${\pmb k}$ and write $g_{\pmb k} = \lambda/|\sqrt{{\pmb k|}}$. We choose a coordinate system so that
  the first 2LS is located at $+\frac{1}{2} {\pmb r}$ and the second at $+\frac{1}{2} {\pmb r}$. Hence,  $\Theta^{(1)}_{\pmb k} = \frac{1}{2} {\pmb k} \cdot {\pmb r} = - \Theta^{(2)}_{\pmb k}$. Then,    $\Sigma^{(1)}(z) = \Sigma^{(2)}(z)  = \Sigma_r(z)$, where
 \begin{eqnarray}
 \Sigma_r(z) =  \frac{\lambda^2}{2\pi^2 r} \left[ [\gamma + \ln(-rz) + \mbox{Cin}(rz)] \sin rz - \mbox{si}(rz)\cos rz\right],
 \end{eqnarray}
 where the
  functions
 $\mbox{Cin}$ and $\mbox{si}$ are   given by Eq. (\ref{CiSi}).

  The  decay constant is
\begin{eqnarray}
\Gamma = \frac{\lambda^2\tilde{\Omega}}{ \pi} (1 \pm \frac{\sin (\tilde{\Omega} r)}{\tilde{\Omega}r}).
\end{eqnarray}

 \bigskip

\noindent 7. We consider  tunneling decays of a particle of mass $m$  through a delta-function potential barrier $V(x) = \eta \delta(x-a)$. The transmission and reflection amplitudes on the real line are,
\begin{eqnarray}
T_k = \frac{1}{1 - \frac{im\eta}{k}} \hspace{1cm} R_k = - \frac{e^{2ika}}{1+ \frac{ik}{m\eta}}.
\end{eqnarray}
For an almost monochromatic initial state with  energy $\frac{k_0^2}{2m}$, the opaque barrier condition $|T_k|<<1$, implies that $\frac{k_0}{m \eta} << 1$. It follows that $g :=  m \eta a >> 1$.
By Eq. (\ref{Gammat}), the dominant contribution to the decay rate is
\begin{eqnarray}
\Gamma = \frac{k_0^3}{2m^3 \eta^2a}, \label{gammatunnel}
\end{eqnarray}
up to corrections of order $g^{-1}$.

As shown in Sec. 5.3.1, we can identify the decay rate by finding the poles $q_n - i \gamma_n$ of $\frac{T_k}{1+R_k}$. To leading order in $g^{-1}$,
\begin{eqnarray}
q_n = \frac{n\pi}{a}( 1 + \frac{1}{2g}) \hspace{1cm} \gamma_n = \frac{n^2\pi^2}{4 a g^2},
\end{eqnarray}
where $n$ is a positive integer. For $n$ such that $q_n \simeq  k_0$, Eq. (\ref{decaytb}) for the decay rate reproduces Eq. (\ref{gammatunnel}).

\end{appendix}

 \end{document}